\documentclass[12pt]{article}
\usepackage{epsfig}
\usepackage{a4}
\usepackage{latexsym}
\usepackage{cite}
\usepackage{amssymb}

\usepackage{pslatex}
\usepackage[latin1]{inputenc}
\usepackage[T1]{fontenc}

\renewcommand{\theequation}{\thesection.\arabic{equation}}

\newcommand{\eqMel}{\raisebox{-0.07cm}{$\stackrel{\rm M}{=} $} }
\newcommand{\eqLL}{\raisebox{-0.07cm}{$\;\stackrel{{\rm LL}}{=}\;$} }
\newcommand{\eqNL}{\raisebox{-0.07cm}{$\;\stackrel{{\rm NL}}{=}\;$} }

\newcommand{\hspn}{{\hspace{-4mm}}}
\newcommand{\hsps}{{\hspace{-3mm}}}
\newcommand{\hspp}{{\hspace{5mm}}}
\newcommand{\beq}{\begin{equation}}
\newcommand{\eeq}{\end{equation}}
\newcommand{\bea}{\begin{eqnarray}}
\newcommand{\eea}{\end{eqnarray}}
\newcommand{\nn}{\nonumber}
\newcommand{\ds}{\displaystyle}
\newcommand{\phm}{\phantom{-}}

\newcommand{\MSb}{$\overline{\mbox{MS}}$}
\newcommand{\as}{\alpha_{\rm s}}
\newcommand{\ar}{a_{\rm s}}
\newcommand{\art}{\tilde{a}_{\rm s}}
\newcommand{\au}{\hat{a}_{\rm s}}
\newcommand{\ra}{\rightarrow}
\newcommand{\ep}{\epsilon}
\newcommand{\eps}{\epsilon^{\:\!2}}

\newcommand{\Ntil}{\widetilde{\!N}}
\newcommand{\GE}{\gamma_{\rm e}}

\begin{document}

\setlength{\parskip}{0.3cm}
\setlength{\baselineskip}{0.54cm}

\def\Ftwo{{F_{\, 2}}}
\def\FL{{F_{\:\! L}}}
\def\F3{{F_{\:\! 3}}}
\def\Qs{{Q^{\, 2}}}
\def\GeV2{{\mbox{GeV}^{\:\!2}}}
\def\x1{{(1 \! - \! x)}}
\def\z#1{{\zeta_{\:\! #1}}}
\def\ca{{C_A}}
\def\cas{{C^{\: 2}_A}}
\def\cat{{C^{\: 3}_A}}
\def\caaf{{C^{\: 4}_A}}
\def\canm{{C^{\: n-1}_A}}
\def\canmm{{C^{\: n-2}_A}}
\def\canmmm{{C^{\: n-3}_A}}
\def\canlm{{C^{\: n-\ell-1}_A}}
\def\canlmm{{C^{\: n-\ell-2}_A}}
\def\canlmmm{{C^{\: n-\ell-3}_A}}
\def\cf{{C_F}}
\def\cfs{{C^{\: 2}_F}}
\def\cft{{C^{\: 3}_F}}
\def\cff{{C^{\: 4}_F}}
\def\cffi{{C^{\: 5}_F}}
\def\cfl{{C^{\: \ell}_F}}
\def\cfn{{C^{\: n}_F}}
\def\cfnm{{C^{\: n-1}_F}}
\def\cfnmm{{C^{\: n-2}_F}}
\def\nf{{n^{}_{\! f}}}
\def\nfs{{n^{\,2}_{\! f}}}
\def\nft{{n^{\,3}_{\! f}}}
\def\caf{{C_{AF}}}
\def\cafs{{C_{AF}^{\: 2}}}
\def\caft{{C_{AF}^{\: 3}}}
\def\caff{{C_{AF}^{\: 4}}}
\def\cafn{{C_{AF}^{\: n}}}
\def\cafnm{{C_{AF}^{\: n-1}}}
\def\cafnmm{{C_{AF}^{\: n-2}}}
\def\dabc2{{d^{\:\!abc}d_{abc}}}
\def\dabcnc{{{d^{\:\!abc}d_{abc}}\over{n_c}}}
\def\dabcna{{{d^{\:\!abc}d_{abc}}\over{n_a}}}
\def\fl11{fl_{11}}
\def\flg11{fl^g_{11}}
\def\fl02{fl_{02}}
\def\b#1{{{\beta}_{#1}}}
\def\bb#1#2{{{\beta}_{#1}^{\,#2}}}

\def\B#1{{{\cal B}_{\:\!#1}}}

%
\begin{titlepage}

\noindent
\hspace*{\fill} LTH 900\\[1mm]   
\hspace*{\fill} December 2010 \\ 
\vspace{2.5cm}
\begin{center}
\Large
{\bf Generalized double-logarithmic large-x resummation \\[1mm]
in inclusive deep-inelastic scattering}\\
\vspace{2cm}
\large
A.A. Almasy, G. Soar and A. Vogt\\
\vspace{1cm}
\normalsize
{\it Department of Mathematical Sciences, University of Liverpool \\
\vspace{0.1cm}
Liverpool L69 3BX, United Kingdom}\\[2.5cm]
\vfill
\large
{\bf Abstract}
\vspace{-0.2cm}
\end{center}
We present all-order results for the highest three large-$x$ logarithms of the
splitting functions $P_{\rm qg}$ and $P_{\rm gq}$ and of the coefficient 
functions $C_{\phi,\rm q}$, $C_{2,\rm g}$ and $C_{L,\rm g}$ for structure
functions in Higgs- and gauge-boson exchange DIS in massless perturbative QCD. 
The corresponding coefficients have been derived by studying the unfactorized partonic structure 
functions in dimensional regularization independently in terms of their 
iterative structure and in terms of the constraints imposed by the functional 
forms of the real- and virtual-emission contributions together with their 
Kinoshita--Lee-Nauenberg cancellations required by the mass-factorization 
theorem. The numerical resummation corrections are small for the splitting 
functions, but partly very large for the coefficient functions. The highest
two (three for $C_{L,\rm g}$) logarithms can be resummed in a closed form in
terms of new special functions recently introduced in the context of the
resummation of the leading logarithms.

\vspace{1cm}
\end{titlepage}

\section{Introduction}
\label{sec:intro}
The splitting functions governing the scale dependence of the parton densities
of hadrons and the coefficient functions for inclusive deep-inelastic 
scattering (DIS) are benchmark quantities of perturbative QCD \cite{PDG2010}. 
At this point these are the only quantities depending on a dimensionless 
scaling variable, the parton momentum fraction and the Bjorken variable (both 
usually denoted by~$x$), for which third-order corrections in the strong 
coupling constant $\as$ are fully known. 
The corresponding three-loop calculations started with DIS sum rules 
\cite{Larin:1991zw,Larin:1991tj,Larin:1991fx} and proceeded via a series of low
integer moments of the splitting functions and the coefficient functions for 
the most important structure functions 
\cite{Larin:1994vu,Larin:1997wd,Retey:2000nq} to the corresponding complete 
calculations of Refs.~\cite{MVV3,MVV4,MVV5,MVV6,MVV10}.
 
Such higher-order calculations do not only improve the numerical accuracy of
the predictions of perturbative QCD but also help to uncover general 
structures, for example in the soft-gluon limit $x \ra 1$. Writing the 
expansion of the splitting functions in the \MSb\ scheme as
\beq
\label{P-exp}
  P_{\:\!ik}^{}(x,\as) \; = \;
  \sum_{n=0}^{\infty}\, \ar^{\,n+1}\, P_{\:\!ik}^{\,(n)}(x)  
\quad \mbox{with} \quad
  \ar \;\equiv\; \frac{\as}{4\pi} \; , 
\eeq
the diagonal (quark-quark and gluon-gluon) splitting functions take the form 
\beq
\label{Pii-xto1}
 P_{\:\! kk}^{\,(n-1)}(x) \; = \;
 A_{\:\! k}^{(n)}\,\x1_+^{-1} 
 \: + \: B_{\:\! k}^{\,(n)}\, \delta \x1 
 \: + \: C_{\:\! k}^{\,(n)}\, \ln \x1
 \: + \: {\cal O}(1) 
\eeq
with the $n$-loop quark and gluon cusp anomalous dimensions related by 
$A_{\,\rm g}^{(n)}/A_{\,\rm q}^{(n)} = \ca/\cf$ at $n \leq 3$, where $\ca$ and 
$\cf$ are the usual SU(N) colour factors with $\ca = 3$ and $\cf = 4/3$ in QCD
\cite{Korchemsky:1989si}. It was not known before Ref.~\cite{DMS05}, inspired 
partly by observations made for the three-loop results in Refs.%
\mbox{\cite{MVV3,MVV4}}, that the third term in Eq.~(\ref{Pii-xto1}) is linear
in $\,\ln \x1$ at all orders $n$, and that its coefficients $C_{\:\!k}^{\,(n)}$
(with $C_{\:\!k}^{\,(1)} = 0\,$) are simple functions of lower-order cusp 
anomalous dimensions.

The form of the off-diagonal (quark-gluon and gluon-quark) splitting functions,
on the other hand, is not stable under higher-order corrections but shows a
double-logarithmic enhancement,
\beq
\label{Pij-xto1}
 P_{\:\! i\neq k}^{\,(n)}(x) \; = \;
 \sum_{\ell=0}^{2n-1}\, D_{\:\! ik}^{\,(n,\ell)}\,\ln^{\,2n-\ell} \x1 
 \: + \: {\cal O}(1) \; . 
\eeq
The terms with $\ell=0$ form the leading-logarithmic (LL) large-$x$
approximation, those with $\ell=1$ the next-to-leading-logarithmic (NLL)
approximation etc. Recently an all-order resummation of the former 
contributions to Eq.~(\ref{Pij-xto1}) has been presented \cite{AV2010}. A main
purpose of this article is to extend those results to the $\ell=2$ 
next-to-next-to-leading logarithmic (NNLL) terms.

The dominant $\x1^{-1}_+$ large-$x$ contributions to the quark coefficient 
functions for gauge-boson exchange structure functions in DIS such as $\Ftwo$ 
and $\F3$ (and the gluon coefficient function for Higgs-exchange structure 
function $F_\phi$ in the heavy-top limit \cite{DGGLcphi,SMVV1}) also show a 
double logarithmic enhancement. 
These terms are resummed to all orders by the soft-gluon exponentiation 
\cite{SGlue1,SGlue2,SGlue3,SGlue4,SGlue5,av00,MVV2,MVV7} which presently fixes 
the coefficients of the highest six logarithms analytically, and the seventh 
term for all numerical purposes since the effect to the presently unknown 
four-loop cusp anomalous dimension can be neglected in this context
\cite{MVV7,Baikov:2006ai,av-dis07}.
A resummation of the highest three $\x1^{0}$ logarithms has been inferred in
Ref.~\cite{MV5} from the properties of the corresponding flavour non-singlet
physical evolution kernels. While those subleading contributions to the 
`diagonal' $O(\as^0)$ coefficient functions are not the main topic of this
article, we will be able to verify the DIS part of those results and fix
the only missing coefficient for the fourth (N$^3$L) logarithms. 

The `off-diagonal' $O(\as^1)$ coefficient functions, such as 
\beq
\label{Cak-exp}
  C_{a, k}(x,\as) \;=\; \sum_{n=1}^{\infty}\, \ar^{\,n}\, 
  c_{a, k}^{\,(n)}(x) \quad \mbox{for} \quad
  a,\, k \:=\: 2,\, {\rm g} ~~\mbox{ or }~~ \phi{,\,\rm q} \; , 
\eeq
receive a double-logarithmic higher-order enhancement as $x \ra 1$ as well,
\beq
\label{Cak-xto1}
 c_{a, k}^{\,(n)}(x) \;=\; \sum_{\ell=0}^{2n-2}\, D_{a,\, k}^{\,(n,\ell)}\,
 \ln^{\,2n-1-\ell} \x1 \: + \: {\cal O}(1) \; . 
\eeq
Also here the $\ell=0\,$ LL coefficients have been determined at all orders $n$
in Ref.~\cite{AV2010}, and also here we will extend those results by deriving 
the corresponding NLL and NNLL results.

Finally we will also address the coefficient functions $C_{L,\rm q}$ and 
$C_{L,\rm g}$ for the longitudinal structure function $\FL$. These quantities
have a perturbative expansion of the form (\ref{Cak-exp}), and are given by
\beq
\label{CLk-xto1}
  c_{L, k}^{\,(n)}(x) \;=\; (1-x)^{\delta_{\:\!k\rm g}} \Bigg(\,
  \sum_{\ell=0}^{2n-3}\, D_{L,\, k}^{\,(n,\ell)}\, \ln^{\,2n-2-\ell} \x1 
  \: + \: {\cal O}(1) \Bigg) \; , \quad k \:=\: {\rm q,\,g} \; , 
\eeq
at large $x$, i.e., they are suppressed by one power of $\x1$ and $\,\ln \x1$
with respect to their counterparts for the structure function $\Ftwo$. The 
coefficients $D_{L,\,\rm q}^{\,(n,\ell)}$ for $\ell = 0,\,1,\,2$ have been 
obtained already in Ref.~\cite{MV3}, again from physical-kernel considerations.
For the gluon coefficient function, however, only the LL coefficients 
$D_{L,\,\rm g}^{\,(n,0)}$ have been determined completely in that article. 
Below we will verify those results and extend also the resummation of 
$C_{L,\rm g}$ to the NNLL terms.

The remainder of this article is organized as follows: In Section 2 we derive
all-order expressions for the (dimensionally regulated) Mellin-$N$ space 
transition functions as far as required for the mass-factorization of the 
structure functions at the level of the dominant and sub-dominant 
contributions discussed above. 
The NNLL resummations of the unfactorized partonic stucture functions $T_{a,k}$
are then derived in Section 3 for the cases (\ref{Cak-exp}), where to NLL
accuract we employ two different methods to drive the same results, and in 
Section 4 for $\FL$.
The partly rather lengthy results of these three sections are then combined, 
and in Sections 5 and 6 we present and discuss the respective resummed 
expressions for the moments of the off-diagonal splitting functions -- which 
receive rather small resummation corrections at relevant values of $N$ --
and the coefficient functions (\ref{Cak-exp}) and $C_{L,\rm g}$ for which 
these corrections are (very) large.
We summarize our findings in Section 7 where we also present a brief outlook to
future extensions and applications of some of the results.
Closed expressions have not been found so far for the third logarithms in 
Eqs.~(\ref{Pij-xto1}) and (\ref{Cak-xto1}). Numerical and symbolic tables of
NNLL coefficients to high orders are therefore finally presented in Appendix A
for the splitting functions and Appendix B for the coefficient functions.

%
\section{Large-$x\,/\:$large-$N$ mass factorization to all orders}
\label{sec:factorization}
\setcounter{equation}{0}
The main part of our calculations is performed after transformation to 
Mellin-$N$ space, 
\beq
\label{Mtrf}
 f(N) \;=\; \int_0^1 \! dx \; x^{\,N-1}\, f(x) \quad \mbox{ or } \quad
 f(N) \;=\; \int_0^1 \! dx \left(\, x^{\,N-1} - 1 \right) f(x)_{+}
 \; , 
\eeq
where the ubiquitous $x$-space Mellin convolutions are reduced to simple 
products. To the accuracy required below, the relations between the large-$x$ 
logarithms and their moment-space counterparts are given~by
\bea
\label{LogsTrf}
  (-1)^{k}\, \Bigg( {\ln^{\,k-1}\x1 \over 1-x} \Bigg)_{\!+} \!
  &\eqMel& {1 \over k}\, \bigg( [S_{1-}(N)]^{\,k} 
     \;+\; {1 \over 2}\, k (k-1) \z2\, [S_{1-}(N)]^{\,k-2} 
  \nn \\[-1mm] & & \mbox{\hspp}
     \;+\; {1 \over 6}\, k (k-1) (k-2) \z3\, [S_{1-}(N)]^{\,k-3} 
     \;+\; O \Big( [S_{1-}(N)]^{\,k-4} \Big) \!\bigg) \; , \nn \\[1mm]
   (-1)^{k}\, \ln^{\,k}\x1 \hspp
  &\eqMel& {1 \over N}\, \bigg( \ln^{\,k} \Ntil
    \;+\; {1 \over 2}\, k (k-1) \zeta_{\,2}\, \ln^{\,k-2} \Ntil
    \;+\; {1 \over 6}\, k (k-1) (k-2) \zeta_{\,3}\, \ln^{\,k-3} \Ntil \quad
  \nn \\ & & \mbox{\hspp}
    \;+\; O \Big( \ln^{\,k-4} \Ntil \Big) \!\bigg) \; , \\[1mm]
  (-1)^{k}\, \x1 \ln^{\,k}\x1 \;
  &\eqMel& {1 \over N^{\,2}}\, \bigg( \ln^{\,k} \Ntil
    \;-\; k\,\ln^{\,k-1} \Ntil 
    \;+\; {1 \over 2}\, k (k-1) \zeta_{\,2}\, \ln^{\,k-2} \Ntil \nn 
    \;+\; O \Big( \ln^{\,k-3} \Ntil \Big) \!\bigg) 
\eea
with $S_{1-}(N) = \ln\,\Ntil - 1/(2N) + O(1/N^{\,2})$ and 
$\:\Ntil \,=\, N e^{\,\GE}$, i.e., $\ln\,\Ntil = \ln N + \GE$ 
with $\GE \simeq 0.577216$.
Here $\eqMel$ indicates that the right-hand-side is the Mellin
transform (\ref{Mtrf}) of the previous expression.  

The primary objects of our resummations are the dimensionally regulated 
unfactorized partonic structure functions or forward Compton amplitudes 
$T_{a,k}$ for the combinations of $a$ and $k$ of Eqs.~(\ref{Cak-exp}) and 
(\ref{CLk-xto1}).
For brevity suppressing all functional dependences on $N$, $\as$ and the 
dimensional offset $\ep$ with $D = 4 - 2\ep$, these quantities
can be factorized as 
\beq
\label{Tdec}
  T_{a, k} \;\;=\;\; \widetilde{C}_{a, i} \: Z_{\,ik}
\:\: .
\eeq
Here the process-dependent $D$-dimensional coefficient functions (Wilson
coefficients) $\widetilde{C}_{a, i}$ include contributions with all 
non-negative powers of $\ep$. The universal transition functions 
(renormalization constants) $Z_{\,ik}$ collect all negative powers of $\ep$ and
are related to the splitting functions in Eq.~(\ref{P-exp}) (or the anomalous
dimensions $\gamma\,$) by
\beq
\label{PofZ}
  - \,\gamma \;\:=\;\: P \;=\; \frac{d\:\! Z }{d\ln \Qs }\; Z^{-1} \; .
\eeq
Here and below we identify, as already in the introduction, the renormalization 
and factorization scale with the physical hard scale $\Qs$ without loss of 
information. Using the $D$-dimensional evolution of the coupling,
\beq
\label{arunD}
  \frac{d\:\! \ar}{d\ln \Qs} \;\:=\;\: -\,\ep\, \ar + \beta (\ar)
\eeq
where $\beta (\ar)$ denotes the usual four-dimensional beta function of QCD,
$\beta (\ar) = - \beta_0 \,\ar^2 + \,\ldots$ with
$\beta_0 = 11/3\:\ca - 2/3\:\nf\,$,
Eq.~(\ref{PofZ}) can be solved for $Z$ order by order in $\as$.

The general elements of $Z^{\,(n)}$ become extremely lengthy at very high 
powers $n$ of~$\ar$. Here, however, we are interested only in the LL, NLL and 
NNLL contributions at order $N^{\,-1}$ for $Z_{\rm qg}$ and $Z_{\rm gq}$ and 
$N^{\,0}$ for $Z_{\rm qq}$ and $Z_{\rm gg}$ (required for Eq.~(\ref{Tdec}) also
in the off-diagonal cases). 
Consequently there can be at most one off-diagonal $N^{\,-1}$ factor per term. 
Moreover $P_{kk}^{(1)}$ in Eq.~(\ref{P-exp}) can enter only (once) at NNLL 
accuracy, and higher-order coefficients $P_{kk}^{(n\geq 2)}$ in 
Eq.~(\ref{P-exp}) are not at all relevant at this level. Finally $\beta_0$ in 
Eq.~(\ref{arunD}) and $\bb02$ only contribute from the NLL and NNLL terms,
respectively, and $\beta_{\:\!n\geq 1}$ would enter only at the next 
logarithmic accuracy. 
All this can be easily read off already from the well-known third-order 
expression for $Z$,
\bea
\label{Z3}
  Z &\!=\!& 1 \:+\: \ar \:{1 \over \ep}\: \gamma^{\,(0)} 
  \:+\: \ar^2 \left( {1 \over 2\:\!\eps}\, 
       \left( \gamma^{\,(0)} - \beta_0 \right) \gamma^{\,(0)}  
       + {1 \over 2\:\!\ep}\: \gamma^{\,(1)} \!\right) 
  \nn \\[0.5mm] & & \;\;\:\mbox{}
  \:+\: \ar^3 \left( {1 \over 6\:\!\ep^3} 
       \left( \gamma^{\,(0)} - \beta_0 \right) 
       \left( \gamma^{\,(0)} - 2\:\!\beta_0 \right) \gamma^{\,(0)} \right.
  \nn \\[0.5mm] & & \left. \mbox{\hspp\hspp}
       + {1 \over 6\:\!\eps} 
       \left[ \left( \gamma^{\,(0)} - 2\:\!\beta_0 \right) 
       \gamma^{\,(1)} + \left( \gamma^{\,(1)} - \beta_1 \right) 
       2\:\!\gamma^{\,(0)} \right] + {1 \over 6\:\!\ep}\: \gamma^{\,(2)} 
       \!\right) 
  \:+\; \ldots
\eea
together with Eqs.~(\ref{Pii-xto1}), (\ref{Pij-xto1}) and (\ref{LogsTrf}) above.

The terms that do contribute to the $a \neq b$ off-diagonal entries of $Z$ at 
the present accuracy can be grouped as follows:
\beq
\label{Zdec}
  Z^{\,(k)}_{ab} \;=\; 
    Z^{\,(k)}_{ab}\Big|_{0} 
  \,+\: Z^{\,(k)}_{ab}\Big|_{\beta_0} 
  \,+\: Z^{\,(k)}_{ab}\Big|_{\beta^{\,2}_{0}} 
  \,+\: Z^{\,(k)}_{ab}\Big|_{\gamma^{\,(1)}\,\gamma^{\,(1)}} 
  \,+\: Z^{\,(k)}_{ab}\Big|_{\gamma^{\,(1)}\,\gamma^{\,(\ell)}} \; .
\eeq
The first term on the right-hand-side collects all contributions with at most
one higher-order anomalous dimension $\gamma^{\,(i\geq 1)}$ but no contribution
from the beta function. It starts at $k=1$ and reads
\bea
\label{Z0}
 Z^{\,(k)}_{ab}\Big|_{0}  &\!=\!&  \frac{1}{k!\, \ep^{\:\!k}}\; \left\{\,
    \sum_{i=0}^{k-1}\: \ep^{\:\!i}\: \sum_{j=0}^{k-1-i}\, 
    \frac{(j+i)!\,}{j!}\, \left(\gamma^{\,(0)}_{aa} \right)^{k-1-i-j}\, 
    \gamma\,^{(i)}_{ab}\, \left(  \gamma\,^{(0)}_{bb} \right)^j \right.
\nn \\
& & \mbox{\hspp\hspp}
   +\: \ep\: \sum_{j=0}^{k-3}\; {1 \over 2}\: (k-j-2)\, (k-j-1)
     \left( \gamma^{\,(0)}_{aa} \right)^j\, \gamma^{\,(0)}_{ab}\: 
     \gamma^{\,(1)}_{bb}\, \left( \gamma\,^{(0)}_{bb} \right)^{k-j-3} 
\nn \\ & & \left.
  \mbox{\hspp\hspp}
   +\: \ep\: \sum_{j=0}^{k-3}\; {1 \over 2}\: (k-j-2) (k+j+1) 
     \left( \gamma^{\,(0)}_{aa} \right)^{k-j-3}\, \gamma^{\,(1)}_{aa}\: 
     \gamma^{\,(0)}_{ab}\, \left( \gamma^{\,(0)}_{bb} \right)^j 
   \right\} \; .
\eea
The first line includes, for $i=0$, the LL expression used in 
Ref.~\cite{AV2010}. The contributions linear in $\beta_0$ contribute from 
$k=2$ and NLL accuracy and are given by
\bea
\label{Zb0}
 Z^{\,(k)}_{ab}\Big|_{\beta_0} &\!\!=\!\!&
 - \,\frac{\beta_{0}}{2}\: \frac{1}{k!\,\ep^{\:\!k}}\: 
 \sum_{i=0}^{k-2}\, \ep^{i}\; \sum_{j=0}^{k-2-i}
 \frac{(i+j)!\,}{j!}\, \left[\, k\,(k-1) - i\,(i+j+1) \right] 
 \left(\gamma^{\,(0)}_{aa}\right)^{k-2-i-j}\, \gamma^{\,(i)}_{ab}\,
 \left(\gamma^{\,(0)}_{bb}\right)^j \: , 
\nn \\[-2mm]
\eea
while the corresponding NNLL $\beta^{\,2}_{0}$ term in Eq.~(\ref{Zdec}) for
$k \geq 3$ is
\bea
\label{Zb02}
 Z^{\,(k)}_{ab}\Big|_{\beta^{\,2}_{0}}  &\!=\!&
   \frac{\beta_{0}^{\,2}}{24}\, \frac{1}{k!\,\ep^{\:\!k}}\: 
   \sum_{i=1}^{k-3}\, \ep^{i}\; \sum_{j=0}^{k-3-i}\, 
   \frac{(j+i)!}{j!}\, \left[\, k\,(k-1)\,(k-2)\,(3\, \*k-1)^{} 
   - 6\,i\,(i+j+1)\,k\,(k-1) \right.
\nn \\[-2mm] & & \left. \mbox{\hspp\hspp\hspp}
   + i\,(3\,i+1)\,(i+j+1)\,(i+j+2)^{} \right] 
   \left(\gamma^{\,(0)}_{aa}\right)^{k-3-i-j}\,\gamma^{\,(i)}_{\,ab}\,
   \left(\gamma^{\,(0)}_{bb}\right)^j \: .
\eea
Finally we distinguish NNLL contributions with $\gamma^{\,(1)}\,\gamma^{\,(1)}$,
where one of the factors is the off-diagonal $O(N^{\,-1})$ entry, and 
contributions with $\gamma\,^{(1)}\,\gamma\,^{(\ell>1)}$, where the latter has 
to be off-diagonal at the present level of accuracy (recall that every term 
includes one off-diagonal anomalous dimension $\gamma_{a \neq b}$).
The former terms contribute from order $\as^{\,4}$ and are given by
\bea
\label{Zg1g1}
 Z^{\,(k)}_{ab}\Big|_{\gamma^{\,(1)}\,\gamma^{\,(1)}} &\!=\!&  
 \frac{1}{k!}\: \ep^{\:\!-k+2}\, \sum_{i=0}^{k-4}\; \sum_{j=0}^{k-4-i}\,  
 \left[ (k-i-2)^2 - 1 - j\,(k-i-1) \right]
\\[1mm] & &
 \mbox{\hspace*{-1.5cm}} \cdot\, \left[ 
 \left( \gamma^{\,(0)}_{aa}\right)^i\,\gamma^{\,(1)}_{ab}\,
 \left( \gamma^{\,(0)}_{bb}\right)^j\,\gamma^{\,(1)}_{bb}\, 
 \left( \gamma^{\,(0)}_{bb}\right)^{k-i-j-4} 
  + \left( \gamma^{\,(0)}_{aa}\right)^i\, \gamma^{\,(1)}_{aa}\,
    \left( \gamma^{\,(0)}_{aa}\right)^j\, \gamma^{\,(1)}_{ab}\, 
    \left( \gamma^{\,(0)}_{bb}\right)^{k-i-j-4\,} \right] 
\nn
\eea
and the corresponding final contribution to Eq.~(\ref{Zdec}) at 
$k \geq 5$ reads
\bea
\label{Zg1gl}
 Z^{\,(k)}_{ab}\Big|_{\gamma^{\,(1)}\,\gamma^{\,(\ell)}} &\!=\!& 
 \frac{1}{k!\,\ep^{k}}\, \sum_{\ell=2}^{k-3}\, \ep^{\:\!\ell+1}\: 
 \left\{ 
 \sum_{i=0}^{k-\ell-3}\; \sum_{j=0}^{k-3-i-\ell}\, 
   (k-i-1)\: \frac{(k-i-j-3)!\,}{(k-i-j-\ell-3)!}
\right.
\nn \\ & & \left. \mbox{\hspp\hspp\hspp\hspp\hspp\hspp} \cdot\, 
 \left( \gamma^{\,(0)}_{\,aa}\right)^i\,\gamma^{\,(1)}_{aa}\, 
 \left( \gamma^{\,(0)}_{\,aa}\right)^j\,\gamma^{\,(\ell)}_{ab}\, 
 \left( \gamma^{\,(0)}_{\,bb}\right)^{k-i-j-\ell-3}
\right.
\nn \\ & & \left. \mbox{\hspp\hspp\hspp\hspp}\;\;
 + \sum_{i=0}^{k-\ell-3}\, \sum_{j=0}^{k-3-i-\ell}\, 
 \frac{(k-1-i)!}{(k-\ell-i-1)!}\: (k-\ell-i-j-2)\, 
\right.
\nn \\ & & \left. \mbox{\hspp\hspp\hspp\hspp\hspp\hspp}\cdot\, 
 \left( \gamma^{\,(0)}_{\,aa}\right)^i\, \gamma^{\,(\ell)}_{ab}\,
 \left( \gamma^{\,(0)}_{\,bb}\right)^j\, \gamma^{\,(1)}_{bb}\, 
 \left( \gamma^{\,(0)}_{\,bb}\,\right)^{k-i-j-\ell-3} \right\} \; . \quad
\eea
The coefficients Eqs.~(\ref{Z0}) -- (\ref{Zg1gl}) have been inferred by 
analyzing the respective first five to seven non-trivial orders $k$ and then 
verified to `all' orders using, as for a large part of our symbolic 
manipulations, the programs {\sc Form} and {\sc TForm} \cite{Form3,TForm}.

The corresponding result for the diagonal entries of the Z-matrices are much 
simpler due to Eq.~(\ref{Pii-xto1}). Including also terms which contribute to 
the next-to-next-to-next-to-leading logarithmic (N$^3$LL) terms suppressed by 
one power of $1/N$, the coefficients at order $\as^{\,k}$ are given by
\bea
\label{Zns}
 Z^{(k)\,}_{aa}  &\!\!=\!\!&
   \frac{\ep^{-k}}{k!}\, \left(\gamma^{\,(0)}_{aa}\,\right)^{k}
 +\; \frac{\ep^{-k+1}}{2\:\!(k-2)!}\,
   \left(\gamma^{\,(0)}_{aa}\,\right)^{k-2} \gamma^{\,(1)}_{aa}
 \;-\; \frac{\beta_{0}}{2}\: \frac{\ep^{-k}}{(k-2)!}\,
   \left(\gamma^{\,(0)}_{aa}\,\right)^{k-1} 
 \nn \\ & & \mbox{}
 +\; \frac{\beta_{0}^{\:\!2}}{24}\: \frac{\ep^{-k}}{(k-3)!}\: (3\:\!k-1)\,
   \left(\gamma^{\,(0)}_{aa}\,\right)^{k-2}
 +\; \frac{\ep^{-k+2}}{3\,(k-3)!}\:
   \left(\gamma^{\,(0)}_{aa}\,\right)^{k-3} \gamma^{\,(2)}_{aa}
 \nn \\ & & \mbox{}
 -\; \frac{\beta_{0}}{12}\: \frac{\ep^{-k+1}}{(k-3)!}\: (3\:\!k-5)\,
   \left(\gamma^{\,(0)}_{aa}\,\right)^{k-3} \gamma^{\,(1)}_{aa}
 \;-\; \frac{\beta_{1}}{3}\:\frac{\ep^{-k+1}}{(k-3)!}\: 
   \left(\gamma\,^{(0)}_{aa}\,\right)^{k-2} 
 \nn \\ & & \mbox{}
 +\; \frac{\ep^{-k+2}}{8\:\!(k-4)!}\: \left(\gamma\,^{(0)}_{aa}\,\right)^{k-4}
   \left(\gamma\,^{(1)}_{aa}\,\right)^{2} 
 +\; \frac{\beta_{0}^{\:\!2}}{48}\: \frac{\ep^{-k+1}}{(k-4)!}\: 
    (k-1)\, (3\:\!k-8)\:
   \left(\gamma^{\,(0)}_{aa}\,\right)^{k-4} \gamma^{\,(1)}_{aa} 
 \nn \\ & & \mbox{}
 -\; \frac{\beta_{0}^{\:\!3}}{48}\:\frac{\ep^{-k}}{(k-4)!}\: k\,(k-1)\:
   \left(\gamma\,^{(0)}_{aa}\,\right)^{k-3} \; .
\eea
Only the first four terms contribute to the $N^{\,0}$ NNLL expression entering 
the off-diagonal $N^{\,-1}$ mass factorization (\ref{Tdec}). Note that, unlike
at order $N^{\,0}$, $P_{aa}^{\,(1)}$ and $P_{aa}^{\,(2)}$ are NLL and N$^3$LL
quantities at order $N^{\,-1}$ due to $C_{\,k}^{\,(1)} = 0\,$ and 
$C_{\,k}^{\,(n\geq 2)} \neq 0\,$ in Eq.~(\ref{Pii-xto1}).

The unfactorized structure functions (\ref{Tdec}) are given by these results 
multiplied by
\beq
\label{Ctilde}
 \widetilde{C}_{a, i} \;=\; 
 \delta_{\,a\:\!\gamma}\, \delta_{\,i\:\!\rm q} \:+\: 
 \delta_{\,a\:\!\phi}\, \delta_{\,i\:\!\rm g}
 \:+\: \sum_{n=1}^{\infty}\, \ar^{\,n}\, \sum_{k=0}^{\infty}\: \ep^{\,k}
 c_{a, i}^{\,(n,k)} \; . 
\eeq
The index $\gamma$ generically represents the gauge-boson exchange structure 
functions (except for $\FL$). $c_{a, i}^{\,(n,0)} \:=\; c_{a, i}^{\,(n)}$ are 
the $n$-th order coefficient functions in Eq.~(\ref{Cak-exp}) for any 
combination of $a$ and $i$. The quantities $c_{a, i}^{\,(n,k)}$ -- usually 
denoted by $a_{a, i}^{\,(n)}$, $b_{a, i}^{\,(n)}$ etc in fixed-order 
calculations -- are enhanced by factors $\ln^{\,k} \Ntil$ with respect to those
four-dimensional coefficient functions.

The calculation of $T_a$ to order $\as^{\,\ell \leq n+1}$ and $\ep^{\,n-\ell}$
(for $\FL$: $\ep^{\,n-\ell+1}$) provides the N$^n$LO (leading-order, 
next-to-leading-order etc) renormalization-group improved fixed-order
approximation to the structure functions $F_a$. 
It is obvious from Eqs.~(\ref{Z3}) -- (\ref{Ctilde}) that a full N$^n$LO result
completely fixes the highest $n\!+\!1$ powers of $1/\ep$ to all orders in $\as$.
An all-order resummation of the splitting functions and coefficient functions 
requires, at the logarithmic accuracy under consideration, an extension of
these results to all powers of $\ep$. 
The flavour-singlet structure functions considered here are fully known at 
N$^2$LO from Refs.~\cite{MVV3,MVV4,MVV5,DGGLcphi,SMVV1} and the earlier 
coefficient-function calculations of Refs.\mbox{\cite{ZvN1,ZvN2,ZvN3,MV00}}. 
Hence a double-logarithmic resummation based on these results can be expected 
to predict up to the highest three logarithms at all higher orders, including 
the corresponding contributions to the three-loop coefficient functions for 
$\Ftwo$ and $F_\phi$ exactly computed in Refs.~\cite{MVV6,SMVV1} and the 
large-$x$ predictions of the four-loop splitting functions in the latter 
article.
 
%
\vspace{-1mm}
\section{Resummation of the unfactorized expressions for $F_2$ and $F_\phi$ }
\vspace{-1mm}
\label{sec:resummation1}
\setcounter{equation}{0}
In this section we derive all-order expressions for the leading $N^{\,-1}$ 
contributions to the off-diagonal amplitudes or unfactorized structure 
functions $T_{2,g}$ and $T_{\phi,q}$ at NNLL accuracy. 
We will apply two approaches: first an iteration of amplitudes generalizing 
the leading logarithmic results of Ref.~\cite{AV2010}, and then an apparently 
new and more rigorous treatment which makes use of only the $D$-dimensional 
structure of the unfactorized structure functions in the large-$x$ limit and 
the KLN cancellations \cite{Kinoshita:1962xx,Lee:1964xx} between its real- and 
virtual-emission contributions.

Both calculations require the corresponding expressions for the $N^{\,0}$ parts
of $Z_{\rm qq}$, $Z_{\rm gg}$, $\widetilde{C}_{2, \rm q}$ and 
$\widetilde{C}_{\phi, \rm g}$ which can be determined from the diagonal 
amplitudes $T_{\,2,\rm q}$ and $T_{\phi,\rm g}$ in the limit governed by the 
soft-gluon exponentiation. These quantities are given by
\beq
\label{Tdiag}
 T_{a,k} \;=\; \exp \left( 
  \au\,       \widetilde{T}_{a,k}^{\,(1)} \:+\: 
  \au^{\,2}\, \widetilde{T}_{a,k}^{\,(2)} \:+\:
  \au^{\,3}\, \widetilde{T}_{a,k}^{\,(3)} \:+\: \ldots \right)
\eeq
with
\beq
\label{Tdexp}
 \widetilde{T}_{a,k}^{\,(n)} \;=\; \sum_{\ell\,=\,-n-1}^\infty \!\ep^{\:\!\ell} 
 \left ( R^{\,(n,\ell)}_{a,k}\: \exp (n\, \ep \ln N) - V^{\,(n,\ell)}_{a,k} \right)
 \; .
\eeq
For the quark case the coefficients entering the highest four logarithms at all
orders in $\as$ and $\ep$~read 
\bea & & 
\label{RV1q}
 R^{\,(1,-2)}_{2,\rm q} \;=\: 4\:, \quad
 R^{\,(1,-1)}_{2,\rm q} \;=\: 3\:, \quad
 R^{\,(1,0)}_{2,\rm q}  \;=\: \phantom{1}7 - 4\,\z2 \:, \quad
 R^{\,(1,1)}_{2,\rm q}  \;=\: 14 - 3\,\z2 - 8\,\z3 \: ,
\nn \\ & &
 V^{\,(1,-2)}_{2,\rm q} \:=\: 4\:, \quad   
 V^{\,(1,-1)}_{2,\rm q} \:=\: 6\:, \quad
 V^{\,(1,0)}_{2,\rm q} \:=\: 16 + 2\,\z2 \:, \quad
 V^{\,(1,1)}_{2,\rm q}  \:=\: 32 - 3\,\z2 - {28 \over 3}\:\z3 \:, \qquad
\eea
\bea
\label{RV2q}
 R^{\,(2,-3)}_{2,\rm q} &\!=\!& V^{\,(2,-3)}_{2,\rm q} \;=\: \b0\:, \quad
 \nn \\[1mm]
 R^{\,(2,-2)}_{2,\rm q} &\!=\!& \bigg(\, {4 \over 3} - 2\,\z2 \bigg)\, \ca
                                 + {19 \over 6}\: \b0 \:, \qquad
 V^{\,(2,-2)}_{2,\rm q} \;=\;  R^{\,(2,-2)}_{2,\rm q} - {3 \over 2}\: \b0 \; ,
 \nn \\[1mm]
 R^{\,(2,-1)}_{2,\rm q}  &\!=\!& 
       \bigg(\, {3 \over 4} - 6\,\z2 + 12\,\z3 \bigg)\, \cf 
     + \bigg(\, {73 \over 18} - 20\,\z3 \bigg)\, \ca 
     + \bigg(\, {373 \over 36} - 3\,\z2 \bigg)\, \b0 \; ,
\nn \\[1mm]
  V^{\,(2,-1)}_{2,\rm q} &\!=\!& 
      \bigg(\, {3 \over 2} - 12\,\z2 + 24\,\z3 \bigg)\, \cf
    - \bigg(\, {41 \over 9} + 26\,\z3 \bigg)\, \ca
    - \bigg(\, {353 \over 18} + \z2 \bigg)\, \b0 \; , 
\\[3mm] 
\label{RV3q}
  R^{\,(3,-4)}_{2,\rm q} &\!=\!& V^{\,(3,-4)}_{2,\rm q} 
                          \;=\: {4 \over 9}\:\bb02 \; , \qquad
  V^{\,(3,-3)}_{2,\rm q} \;=\:  R^{\,(3,-3)}_{2,\rm q} + \bb02 \; , 
 \nn \\[1mm]
  R^{\,(3,-3)}_{2,\rm q} &\!=\!&
    - {14 \over 9}\:\cas 
    - {22 \over 9}\:\cf \ca
    + {2 \over 3}\:\cf \b0
    + \bigg(\, {62 \over 27} - {16 \over 9}\:\z2 \bigg)\, \ca \b0
    + {67 \over 27}\:\bb02 
\eea
and
\beq
\label{RV4q}
  R^{\,(4,-5)}_{2,\rm q} \;=\; V^{\,(4,-5)}_{2,\rm q} \;=\; \frac{1}{4}\:\bb03
  \; , \qquad
\eeq
where we have suppressed an obvious overall factor of $\cf$ and expressed the
dependence of the number $\nf$ of effectively massless flavours in terms of
$\b0$.
To N$^3$LL accuracy these results are converted to the renormalized coupling 
used elsewhere in this article via 
\beq
\label{asren}
 \au \;=\; \ar \:-\: {\b0 \over \ep}\; \ar^2 
  \:+\: \bigg(\, {\bb02 \over \eps} - {\b1 \over 2\:\!\ep}\,\bigg)\: \ar^3
  \:+\: {\bb03 \over \ep^3}\; \ar^4 \:+\: \ldots \; .
\eeq
The corresponding gluonic coefficients are required only to NNLL accuracy 
here and read
\bea 
\label{RV1g}
 R^{\,(1,-2)}_{\phi,\rm g} &\!=\!& 4\:\!\ca\:, \quad
 R^{\,(1,-1)}_{\phi,\rm g} \;=\: \b0\:, \quad
 R^{\,(1,0)}_{\phi,\rm g}  \;=\: 
  \bigg(\, {4 \over 3} - 4\:\!\z2 \bigg) \ca + {5 \over 3}\:\b0 \; , 
 \nn \\
 V^{\,(1,-2)}_{\phi,\rm g} &\!=\!& 4\:\!\ca\:, \quad
 V^{\,(1,-1)}_{\phi,\rm g} \:=\: 0 \:, \quad \;\;
 V^{\,(1,0)}_{\phi,\rm g} \:=\: 2\,\z2\,\ca \:, \quad
 \\[2mm]
\label{RV2g}
 R^{\,(2,-3)}_{\phi,\rm g} &\!=\!& 
 V^{\,(2,-3)}_{\phi,\rm g} \;=\: \ca\b0\:, \quad
 \nn \\[1mm]
 R^{\,(2,-2)}_{\phi,\rm g} &\!=\!& \bigg(\, {4 \over 3} - 2\,\z2 \bigg)\, \cas
         + {5 \over 3}\: \ca\b0 + {1 \over 2}\:\bb02 \; , \quad
 V^{\,(2,-2)}_{\phi,\rm g} \;=\;  
 R^{\,(2,-2)}_{\phi,\rm g} - {1 \over 2}\: \bb02 \; ,
 \\[2mm]
\label{RV3g}
 R^{\,(3,-4)}_{\phi,\rm g} &\!=\!& V^{\,(3,-4)}_{\phi,\rm g}
                          \;=\: {4 \over 9}\:\ca\bb02 \; . \qquad
\eea
After the transformation to the renormalized coupling, Eqs.~(\ref{Tdiag}) and
(\ref{Tdexp}) for $T_{\phi,\rm g}$ need to be multiplied by the
renormalization constant of $G^{\,\mu\nu} G_{\mu\nu}$ \cite{ZGG1,ZGG2}, 
\beq
\label{GGren}
  1 \:-\: 2\:\!\b0\, \ep^{-1}\: \ar^2 
    \:+\: 3\:\!\bb02\, \ep^{-2}\: \ar^3 \:+\: \ldots \; .
\eeq
  
\vspace*{-3mm}
Note that there is no new physics content in Eqs.~(\ref{Tdiag}) -- (\ref{RV3g})
which represent the Mellin transforms of the decomposition of $T_{2,\rm q}$
and $T_{\phi,\rm g}$ used in Refs.~\cite{MVV8,MVV9} and its obvious 
higher- order generalizations, cf.~Ref.~\cite{MV4}, recast in an exponential
all-order form.
Recall also, from the same references, that of the $\ell$ coefficients 
$R^{\,(1,\:\!\ell-3)}$, $R^{\,(2,\:\!\ell-5)} \;\ldots\; 
R^{\,(n,\:\!\ell-2n-1)}$ relevant for the $\ell$-th logarithm at order $n$ only 
one combination is not fixed by lower-order information, and that the same 
holds for their virtual-correction counterparts $V^{\,(1,\:\!\ell-3)}$, 
$V^{\,(2,\:\!\ell-5)} \;\ldots\; V^{\,(n,\:\!\ell-2n-1)}$. 

We are now ready to address the resummation of the off-diagonal amplitudes.
Using the particularly simply colour structure of the identical 
leading-logarithmic contributions to $T_{\phi,\rm q}/\cf$ and 
$T_{2,\rm g}/\nf$, their LL resummation has been inferred in 
Ref.~\cite{AV2010} to which the reader is referred for a detailed discussion.
The result can be written as
\beq
\label{ToffLL}
  T_{a,k}^{\,(n)} \;\eqLL\; {1 \over n}\: T_{a,k}^{\,(1)}\: \sum_{i=0}^{n-1}\: 
  \Big(\! \begin{array}{c} n\!-\!1 \\[-0.5mm] i \end{array} \!\Big)^{-1}
  T_{\phi,\rm g}^{\,(i)} \: T_{2,\rm q}^{\,(n-i-1)} 
\eeq
where, of course, only the respective first terms of 
\bea
\label{Toff1}
  T_{\phi,\rm q}^{\,(1)}  &\!=\!& 
  - \,{\cf \over N\:\!\ep}\: \exp\, (\ep \ln N)\:
  \left( 2 \:-\: \:\ep\: \:-\: ( 3 - 2\,\z2)\,\eps \:+\: \ldots \right) \; ,
 \nn \\[1mm] 
  T_{\:\!2,\rm g}^{\,(1)} &\!=\!& 
  - \,{\nf \over N\:\!\ep}\: \exp\, (\ep \ln N)\:
  \left( 2 \:- 2\,\ep -\: ( 6 - 2\,\z2)\,\eps \:+\: \ldots \right)
\eea
are required.
The corresponding expressions for $T_{2,\rm q}$ and $T_{\phi,\rm g}$ can be 
read off from Eqs.~(\ref{Tdiag}), (\ref{Tdexp}), (\ref{RV1q}) and (\ref{RV1g})
above.

Eq.~(\ref{ToffLL}) can be generalized to next-to-leading logarithmic accuracy,
where running-coupling effects enter for the first time, using the natural 
ansatz
\beq
 T_{a,k}^{\,(n)} \;\eqNL\;
 { 1 \over n } \: T_{a,k}^{\,(1)}
 \!\left\{ \,\sum_{i=0}^{n-1}\, f_{a,k}(n,i,\ep)\,
  T_{\phi,\rm g}^{\,(i)} \, T_{2,\rm q}^{\,(n-i-1)} 
  \;-\; {\beta_0 \over \ep}\: \sum_{i=0}^{n-2}\, g_{a,k}(n,i)\,
  T_{\phi,\rm g}^{\,(i)} \, T_{2,\rm q}^{\,(n-i-2)} \!\right\} 
\label{eq:NLLit}
\eeq
where $f_{a,k}(n,i,\ep)$ are linear functions of $\ep$. It is not a priori 
clear that such a simple ansatz is compatible with the infinite number of 
constraints provided by the NLL coefficients of the four highest powers of 
$1/\ep$ at all orders $n$ in $\as$ which are provided by the results of
Refs.\mbox{\cite{MVV3,MVV4,MVV6,SMVV1}} including the large-$x$ results for the 
four-loop splitting functions $P_{\rm qg}^{(3)}$ and $P_{\rm gq}^{(3)}$ in 
Eq.~(\ref{P-exp}). However, all these constraints can indeed be fulfilled, and
the resulting coefficients are given by
\bea
  f_{\phi,\rm q}(n,i,\ep) &\!=\!& 
    \Big(\! \begin{array}{c} n\!-\!1 \\[-0.5mm] i \end{array} \!\Big)^{-1}
    \,\left[ \, 1 \:+\: \ep\, 
      \left( {\beta_0 \over 8 \ca}\: (i+1)(n-i)\: \theta_{\,i1}
      - {3 \over 2}\: ( 1 - n\:\! \delta_{\,i\,0} ) \right) \right] \; ,
\nn \\[1.5mm]
  f_{2,\rm g}(n,i,\ep) &\!=\!& 
    \Big(\! \begin{array}{c} n\!-\!1 \\[-0.5mm] i \end{array} \!\Big)^{-1}
    \,\left[ \, 1 \:+\: \ep\, 
      \left( {\beta_0 \over 8 \ca}\: i(n-i-3)\,
      + {1 \over 2}\, (3i+1 - n\,\theta_{\,n2}\, \delta_{\,i\;n\!-\!1} ) 
    \right) \right] \quad
\label{eq:f2g}
\eea
and
\beq
  g_{\phi,\rm q}^{}(n,i) \;=\; g_{2,\rm g}^{}(n,i) \;=\;  
     \Big(\! \begin{array}{c} n \\[-0.5mm] i+\!1  \end{array} \!\Big)^{-1} 
\label{eq:gphiq2g}
\eeq
with $\theta_{kj} = 1$ for $k\geq j$ and $\theta_{kj} = 0$ else.
Eqs.~(\ref{eq:NLLit}) -- (\ref{eq:gphiq2g}) together with their diagonal
counterparts (\ref{Tdiag}), keeping the $\ep^{-2}$ and $\ep^{-1}$ contributions
to $\widetilde{T}_{a,k}^{\,(1)}$ and the $\ep^{-3}$ terms of 
$\widetilde{T}_{a,k}^{\,(2)}$ in Eq.~(\ref{Tdexp}), facilitate the extension 
of the all-order mass factorization to the next-to-leading logarithms.

An extension of Eq.~(\ref{eq:NLLit}) to the the third logarithms can be
expected to become much more cumbersome, requiring at least $\eps$ corrections 
to Eqs.~(\ref{eq:f2g}), $\ep$ corrections to Eqs.\ (\ref{eq:gphiq2g}) and a new
$\bb02$ contribution, but presumably also terms respectively involving 
$T_{2,\rm g}^{\,(2)}$ or $T_{\phi,\rm q}^{\,(2)}$. Instead of pursuing this 
approach, we now switch to our second method for the resummation of 
$T_{2,\rm g}$ and $T_{\phi,\rm q}$.

For this purpose we consider the calculation of $T_{a,k}$ via suitably 
projected gauge-boson parton cross sections as performed at two loops in 
Refs.~\cite{ZvN1,ZvN2,ZvN3}. The maximal ($\:\!2 \ra n+1$ particles) phase 
space for these processes at order $\as^{\,n}$ is 
\cite{Matsuura:1987wt,Matsuura:1989sm}
\beq
\label{PSn}
 \x1^{-1-n\,\ep}\: {\int_0^1} d(3\:\!n\!-\!2 \mbox{ other variables})
 \: f(x,\,\ldots) \; ,
\eeq
where one trivial azimuthal integration has not been counted. 
The integrals for the $n$-th order purely real (tree graph) contributions 
$T_{a,j}^{\:\!(n)\rm R}$ do not lead to any further factors $\x1^{-\ep}$, hence
their expansion around $x=1$ can be written as
\beq
\label{Treal}
  T_{a,j}^{\,(n)\rm R} \;=\; \x1^{-1-n\,\ep}\, \sum_{\xi=0}^\infty\, 
  \x1^\xi \:{1 \over \ep^{\,2n-1} } \:
  \Big\{
           R_{a,j,\,\xi}^{\,(n)\rm LL} \:+\: 
    \ep\,  R_{a,j,\,\xi}^{\,(n)\rm NLL} \:+\: 
    \eps\, R_{a,j,\,\xi}^{\,(n)\rm NNL} \:+\: \ldots 
  \Big\} \; .
\eeq
The mixed contributions ($\:\!2 \ra r+1$ particles with $n-r \geq 1$ loops) 
include up to $n-r$ additional factors of $\x1^{-\ep}$ from the loop
integrals on top the phase-space factor, leading to
\bea
\label{Tmixed}
 T_{a,j}^{\,(n)\rm M} &\!=\!& \sum_{\ell=r}^{n}\, \x1^{-1-\ell\,\ep} 
 \sum_{\xi=0}^\infty\, \x1^\xi\: {1 \over \ep^{2n-1} }\: \cdot
\nn \\[0.5mm] & & \mbox{\hspp} \cdot
  \Big\{
           M_{a,j,\ell,\,\xi}^{\,(n)\rm LL} \:+\: 
    \ep\,  M_{a,j,\ell,\,\xi}^{\,(n)\rm NLL} \:+\: 
    \eps\, M_{a,j,\ell,\,\xi}^{\,(n)\rm NNL} \:+\: \ldots
  \Big\} \; .
\eea
Finally the diagonal cases, where terms with $\xi=0$ are present in 
Eqs.~(\ref{Treal}) and (\ref{Tmixed}), also receive purely virtual 
contributions $T_{a,j}^{\,(n)\rm V}$ given by the $\gamma^{\,\ast\!}qq$ and 
$\phi gg$ form factors which are known to an amply sufficient accuracy
\cite{MVV8,MVV9,FFnew1,FFnew2,FFnew3,FFnew4}
\beq
\label{Tvirt}
  T_{a,j}^{\,(n)\rm V} \;=\; \delta \x1 \:{1 \over \ep^{\,2n} }\:
  \Big\{
          V_{a,j}^{\,(n)\rm LL} \:+\: 
   \ep\,  V_{a,j}^{\,(n)\rm NLL} \:+\: 
   \eps\, V_{a,j}^{\,(n)\rm NNL} \:+\: \ldots
  \Big\} \; . 
\eeq
Note that for $\xi=0$ the $\x1$ factors in Eqs.~(\ref{Treal}) and 
(\ref{Tmixed}) are $D$-dimensional +-distributions which include a factor
$\ep^{-1} \delta \x1$ after expansion in $\ep$.

The partonic cross sections $T_{a,k}^{\,(n)}$ at order $\as^{\,n}$ do not 
include any poles higher than $\ep^{-n}$, i.e., 
\bea
\label{Ttotal}
 T_{a,k}^{\,(n)} &\!=\!& T_{a,k}^{\,(n)\rm R} \,+\: T_{a,k}^{\,(n)\rm M}
 \,+\: ( \delta_{\,a\:\!\gamma}\, \delta_{\,i\:\!\rm q} \:+\:
         \delta_{\,a\:\!\phi}\, \delta_{\,i\:\!\rm g} )\, 
         T_{a,k}^{\:\!(n)\rm V} 
 \nn \\[1mm]
 &\!=\!& {1 \over \ep^{\,n}} \, 
 \Big\{ T_{a,j}^{(n)0} \,+\, \ep\:\! T_{a,j}^{(n)1} 
        \,+\, \ep\:\! T_{a,j}^{(n)2} \,+\, \dots \Big\}
\eea
(recall Eq.~(\ref{Ctilde}) concerning the index $\gamma\,$). Hence there are 
$\,n-1-m\,$ KLN relations between the $n$ N$^{\:\!m\:\!}$LL coefficients in 
Eqs.~(\ref{Treal}) and (\ref{Tmixed}). Since a full N$^{\:\!k\:\!}$LO 
calculation provides, as discussed above, the first $k+1$ non-trivial powers of 
$\ep^{-1}$, it leads to a total of $\,n-m+k\,$ relations. Consequently the 
coefficients up to the N$^{\:\!k\:\!}$LL terms are fixed (and those for $m<k$ 
over-constrained) in terms of the N$^{\:\!k\:\!}$LO results by nothing but the 
above $D$-dimensional structure and the mass-factorization formula (\ref{Tdec})
which is guaranteed for structure functions in DIS by the operator-product 
expansion \cite{OPE}, see also, e.g., Refs.~\cite{BurasRMP,ReyaPR}.

Note that this resummation is far less predictive than the soft-gluon
exponentiation \cite{SGlue1,SGlue2,SGlue3,SGlue4,SGlue5} of the $\:\!\x1^{-1}
\!/ N^{\,0}$ terms of $C_{2,\rm q}$ and $C_{\phi,\rm g}$ which involves an 
additional factorization in the threshold limit.  Therefore, as mentioned below 
Eq.~(\ref{GGren}), only one of the $n$ N$^m$LL coefficients for $\xi=0$ is
actually independent. Hence a N$^n$LO calculation in this case implies a
N$^n$LL exponentiation which fixes the highest $2\:\!n+1$ logarithms at all
orders in $\as$.

Now we switch back to Mellin space and apply the above results to 
$T_{\,2,\rm g}$ and $T_{\phi,\rm q}$. 
The $n$-th order contributions to these quantities can be written as
\beq
\label{T2pABC}
  T^{\,(n)}_{a,k}(N) \;=\; 
  \frac{1}{N\,\ep^{\:\!2n-1}}\: \sum_{i=0}^{n-1}\, 
  \left( 
         A^{\,(n,i)}_{a,k} \:+\: 
  \ep\,  B^{\,(n,i)}_{a,k} \:+\:  
  \eps\, C^{\,(n,i)}_{a,k} \:+\: \ldots\,
  \right)\, \exp (\ep\, (n-i)\, \ln N) \; .
\eeq
As discussed below Eq.~(\ref{Ttotal}), only one of the LL coefficients 
$A^{\,(n,i)}_{a,k}$ is independent for each value of $n$ and $a$, $k$.
Choosing $\,i=0$ for that, the KLN constraints on the other coefficients read
\beq
\label{A-KLN}
   A^{\,(n,i)}_{a,k} \;=\;  (-1)^{i}\, \* 
   \left( \begin{array}{cc} \!\!  n - 1 \!\! \\ i  \end{array} \right) 
   \, A^{\,(n,0)}_{a,k} \; .
\eeq
The $\,i=0\,$ coefficients are found to be
\beq
\label{A-LO}
  {1 \over \nf}\: A^{\,(n,0)}_{2,\rm g} \;=\;
  {1 \over \cf}\: A^{\,(n,0)}_{\phi,\rm q} \;=\; \mbox{}
  - 2^{2\:\! n - 1}\, \frac{1}{n!}\: 
  \sum_{\ell=0}^{n-1}\, C_F^{\,\ell}\, C_A^{\,n-\ell-1} \; .
\eeq
 
At NLL level two coefficients in the sum in Eq.~(\ref{T2pABC}) are not fixed
by the KLN cancellations. 
In terms of our choice $\,i=0$ and $\,i=1$ the remaining coefficients are 
given by
\beq
\label{B-KLN}
  B^{\,(n,i+1)}_{a,k} \;=\; (-1)^{i}\, \left[ 
  \left( \begin{array}{cc} \!\! n - 2 \!\! \\  i \end{array} \right) \,
  B^{\,(n,1)}_{a,k} \:+\: i 
  \left( \begin{array}{cc} \!\! n - 1 \!\! \\ 
                           \!\!  i + 1 \!\! \end{array} \right) 
  \, B^{\,(n,0)}_{a,k} \right] \; .
\eeq
The all-order results for the respective two coefficients determined by the 
NLO results are 
\bea
\label{B2g0}
 B^{\,(n,0)}_{2,\rm g}  &\!=\!&  
-\,\frac{4^{n-2}}{6n!}\,\nf\:
\Bigg\{\,
        \sum_{\ell=1}^{n-2}\cfl\canlm \left(
          11 n^2
          + 39n
          + 22 - 14\ell
        \right) \nn \\[1mm]
& & \quad\quad
                +\,36\,\cfnm(n+1)\,\theta_{n2}\
                +\ \canm\left(
                        11n^2
                        + 15n
                        + 22
                \right)\nn \\
& & \quad\quad
                -\,2\nf\sum_{\ell=0}^{n-2} \cfl\canlmm\left(
                        n^2
                        - 3n
                        + 2(\ell + 1)
                \right)
\Bigg\} \:\: ,
\\[3mm]
\label{B2g1}
 B^{\,(n,1)}_{2,\rm g} &\!=\!& \frac{4^{n-2}}{6n!}\,\nf\:
\Bigg\{\,
        \sum_{\ell=1}^{n-2}\cfl\canlm \left(
                11n^3
                + 50\,n^2
                - \left(9 + 14(\ell - 1)\right)n
                - 66 + 6\ell
        \right) \nn \\[1mm]
& & \quad\quad
                +\,36\,\cfnm(n+2)(n-1)\
                +\ \canm\left(
                        11n^3
                        + 26n^2
                        + 29n
                        - 66
                \right)\nn \\
& & \quad\quad
                -\,2\nf\sum_{\ell=0}^{n-2}\cfl\canlmm\left(
                        n^3
                        - 2n^2
                        + (7 + 2\ell)n
                        - 6(\ell + 1)
                \right)
\Bigg\}
\eea
and 
\bea
\label{Bpq0}
 B^{\,(n,0)}_{\phi,\rm q}  &\!=\!& -\,\frac{4^{n-2}}{6n!}\,\cf\:
\Bigg\{\,
        \sum_{\ell=0}^{n-2}\cfl\canlm\left(
                11n^2
                + 33n
                - 70 + 14\ell
        \right)\nn \\[-1mm]
& & \quad
        +\,12\,\cfnm(9n - 7)\
        -\ 2\nf\sum_{\ell=0}^{n-2}\cfl\canlmm(
                n^2
                + 3n
                - 2(\ell + 1)
        )
\Bigg\} \:\: ,
\\[3mm]
\label{Bpq1}
 B^{\,(n,1)}_{\phi,\rm q}  &\!=\!& \frac{4^{n-2}}{6n!}\,\cf\:
\Bigg\{\,
        \sum_{\ell=0}^{n-2}\cfl\canlm\left(
                11n^3
                + 44n^2
                + \left(14\ell - 81\right)n
                + 26 - 22\ell
        \right)\nn \\[1mm]
& & \quad
        +\,12\,\cfnm(
                9n^2
                - 13n
                + 4
        )\nn \\
& & \quad
        -\,2\nf\sum_{\ell=0}^{n-2}\cfl\canlmm(
                n^3
                + 4n^2
                - (3 + 2\ell)n
                - 2(\ell + 1)
        )
\Bigg\} \:\: .
\eea
Similar to Eqs.~(\ref{Z0}) -- (\ref{Zg1gl}), we have first inferred these
results by analyzing a couple of orders (enough to over-constrain the 
numerator polynomials) and then verified them to `all' orders. 
Eqs.~(\ref{T2pABC}) -- (\ref{Bpq1}) lead to the same results as 
Eqs.~(\ref{Toff1}) -- (\ref{eq:gphiq2g}) above.

At NNLL accuracy, finally, all but three coefficients (chosen as $i=0,\,1,\,2$)
in the sum in Eq.~(\ref{T2pABC}) are specified by the vanishing of poles 
higher that $\ep^{-n}$ in $T^{\,(n)}_{2,\rm g}$ and $T^{\,(n)}_{\phi,\rm q}$.
These coefficients can be written as
\beq
\label{C-KLN}
  C^{\,(n,i+2)}_{a,k} \;=\; (-1)^{i}\, \left[
  \left( \begin{array}{cc} \!\! n - 3 \!\! \\  i \end{array} \right) \,
  C^{\,(n,2)}_{a,k} \:+\: i
  \left( \begin{array}{cc} \!\! n - 2 \!\! \\
                           \!\! i + 1 \!\! \end{array} \right) \, 
  C ^{\,(n,1)}_{a,k} \:+\: {1 \over 2}\: i\, (i+1)
  \left( \begin{array}{cc} \!\! n - 1 \!\! \\ 
                           \!\! i + 2 \!\! \end{array} \right) \, 
  C ^{\,(n,0)}_{a,k} \right] \; .
\eeq
The general expressions for the three independent coefficient at rather
lengthy. However, especially since we will not be able to express the NNLL
mass-factorized results in a closed form, they are presented here nevertheless
in order to assist future research by others. The coefficients for 
$T_{2,\rm g}$ are
\bea
\lefteqn{ 
  C^{\,(n,0)}_{2,\rm g} \ =\ -\,\frac{4^{n-2}}{6n!}\,\nf\:
  \Bigg\{ 
  	\canm\left(
		76n^2 
		+ 60n 
		+ 8
	\right)\ 
	-\ \canm\z2\left(
		30n^2 
		+ 30n 
		- 12\right)
	}\nn \\
& & +\,300\,\cf\delta_{n2}\ 
	-\ 48\cfnm\z2\,n\,\theta_{n2}\ 
	+\ \b0\canmm\left(
		3n^3 
		- 4n^2 
		- 9n 
		+ 10
	\right)\theta_{n2} \nn\\[1mm]
& & +\,\cfnm\bigg(\,
		\frac{27}{2}n^2 
		+ \frac{195}{2}n 
		+ 111\!
	\bigg)\theta_{n3}\ 
	+\ \cf\canmm\left(
		28n^2 
		+ 132n 
		- 68
	\right)\theta_{n3}\nn\\
& & +\,\sum_{\ell=0}^{n-3} \bb02\,\cfl\canlmmm\bigg[\,
		\frac{3}{32}n^4 
		- \frac{29}{48}n^3 
		+ \bigg(\,\frac{3}{8}\ell + \frac{41}{32}\,\bigg)n^2 
		- \bigg(\,\frac{5}{8}\ell + \frac{49}{48}\,\bigg)n 
		+ \frac{1}{8}(\ell^2 + 3\ell + 2)
	\bigg] \theta_{n3}\nn\\
& & +\,\sum_{\ell=1}^{n-2} \b0\cfl\canlmm\bigg[\,
		\frac{9}{2}n^3 
		- \bigg(\,\frac{9}{4}\ell + \frac{17}{2}\,\bigg)n^2 
		+ \bigg(\,\frac{45}{4}\ell - 6\bigg)n 
		- \frac{9}{4}\ell^2 + \frac{31}{4}\ell + 10
	\bigg] \theta_{n3}\nn\\
& & -\,\sum_{\ell=1}^{n-2} \cfl\canlm\z2\left[
		6n^2 
		+ 54n 
		- 12(\ell + 1)
	\right] \theta_{n3}\nn\\
& & +\,\sum_{\ell=2}^{n-2} \cfl\canlm\bigg[
		58n^2 
		- (54\ell - 156)n 
		+ \frac{27}{2}\ell^2 - \frac{179}{2}\ell + 8
	\bigg]\theta_{n4}
\Bigg\} \:\: ,
\eea
\bea
\lefteqn{ C^{\,(n,1)}_{2,\rm g} \ =\ \frac{4^{n-2}}{6n!}\,\nf\,
\Bigg\{
	\canm\left(
		76n^3 
		- 24n^2 
		- 52n
	\right)\ 
	-\ \canm\z2\left(
		30n^3 
		- 12n^2 
		- 6 n 
		- 12
	\right) }\nn \\ 
& & +\,456\,\cf\delta_{n2}\ 
	-\ \cfnm\z2\left(
		48n^2 
		- 72n 
		- 24
	\right)\theta_{n2}\ 
	+\ \b0\canmm\left(
		3n^4 
		- n^3 
		- 21n^2 
		+ 49n 
		- 30
	\right)\theta_{n2} \nn\\[1mm]
& & +\,\cfnm\left(
		\frac{27}{2}n^3 
		+ 111n^2 
		+ \frac{189}{2}n 
		- 225
	\right)\theta_{n3}\ 
	+\ \cf\canmm\left(
		28n^3 
		+ 96n^2 
		- 152n 
		+ 144\right)\theta_{n3} \nn\\
& & +\,\sum_{\ell=0}^{n-3} \bb02\,\cfl\canlmmm\bigg[\,
		\frac{3}{32}n^5 
		- \frac{31}{96}n^4 
		+ \bigg(\,\frac{3}{8}\ell + \frac{13}{96}\,\bigg)n^3 
		- \bigg(\ell + \frac{113}{96}\,\bigg)n^2
	\nn\\
& & \quad\quad
		+ \bigg(\,\frac{1}{8}\ell^2 + 4\ell + \frac{217}{48}\,\bigg)n 
		- \frac{13}{8}(\ell^2 + 3\ell + 2)
	\bigg] \theta_{n3}\nn\\
& &	+\,\sum_{\ell=1}^{n-2} \b0\cfl\canlmm\bigg[\,
		\frac{9}{2}n^4 
		- \bigg(\,\frac{9}{4}\ell + 4\bigg)n^3 
		+\bigg(\,\frac{45}{4}\ell - \frac{33}{2}\,\bigg)n^2 
		-\bigg(\,\frac{9}{4}\ell^2 + \frac{113}{4}\ell - 46\bigg)n
	\nn\\
& & \quad\quad 
		+ \frac{27}{2}\ell^2 - \frac{33}{2}\ell - 30
	\bigg] \theta_{n3}\nn\\
& & +\,\sum_{\ell=2}^{n-2} \cfl\canlm\bigg[
		58n^3 
		- (54\ell - 90)n^2 
		+ \bigg(\,\frac{27}{2}\ell^2 + \frac{37}{2}\ell - 154\bigg)n 
		- \frac{81}{2}\ell^2 + \frac{369}{2}\ell
	\bigg]\theta_{n4}
        \nn\\
& &	-\,\sum_{\ell=1}^{n-2} \cfl\canlm\z2\left[
		6n^3 
		+ 36n^2 
		- (12\ell + 78)n 
		+ 12(\ell - 1)
	\right]\theta_{n3} 
\Bigg\} \:\: ,
\eea
\bea
\lefteqn{ C^{\,(n,2)}_{2,\rm g} \ =\ -\,\frac{4^{n-2}}{6n!}\,\nf\,
\Bigg\{
	\canm\left(
		38n^4 
		- 92n^3 
		+ 6n^2 
		+ 56n 
		- 8
	\right)} \nn\\ 
& & -\,\canm\z2\left(
		15n^4 
		- 42n^3 
		+ 39n^2 
		- 48n 
		+ 36
	\right)
    -\,\cfnm\z2\left(
		24n^3 
		- 96n^2 
		+ 72n 
		+ 48
	\right)\theta_{n2}  \qquad\nn\\[1mm] 
& & +\,\b0\canmm\bigg[\,
		\frac{3}{2}n^5 
		- \frac{1}{2}n^4 
		- \frac{47}{2}n^3 
		+ \frac{153}{2}n^2 
		- 104n 
		+ 50
	\bigg] \theta_{n2}
    +\,\cfnm\bigg[\,
		\frac{27}{4}n^4 
		+ \frac{111}{2}n^3 
\nn\\[1mm] & & \quad
		- \frac{147}{4}n^2 
		- \frac{771}{2}n 
		+ 366
	\bigg]\theta_{n3}
    +\,\cf\canmm\bigg[\,
		14n^4 
		+ 16n^3 
		- 140n^2 
		+ 214n 
		- 220
	\bigg] \theta_{n3}\nn\\
& &	+\,\sum_{\ell=0}^{n-3} \bb02\,\cfl\canlmmm\bigg[\,
		\frac{3}{64}n^6 
		- \frac{13}{192}n^5 
		+\bigg(\,\frac{3}{16}\ell - \frac{31}{64}\,\bigg)n^4 
		-\bigg(\,\frac{7}{8}\ell - \frac{353}{192}\,\bigg)n^3
	\nn\\
& & \quad\quad
		+\bigg(\,\frac{1}{16}\ell^2 + 3\ell + \frac{33}{16}\,\bigg)n^2 
		-\bigg(\,\frac{27}{16}\ell^2 + \frac{211}{16}\ell + 
                    \frac{607}{48}\,\bigg)n 
		+ \frac{37}{8}(\ell^2 + 3\ell + 2)
	\bigg] \theta_{n3}\nn\\
& & +\,\sum_{\ell=1}^{n-2} \b0\cfl\canlmm\bigg[\,
		\frac{9}{4}n^5 
		- \left(\frac{9}{8}\ell + 2\right)n^4 
		+ \left(\frac{27}{4}\ell - \frac{91}{4}\right)n^3
	\nn\\
& & \quad\quad
		-\bigg(\,\frac{9}{8}\ell^2 + 31\ell - \frac{147}{2}\,\bigg)n^2 
		+\bigg(\,\frac{117}{8}\ell^2 + \frac{359}{8}\ell - 101\bigg)n 
		- \frac{135}{4}\ell^2 + \frac{65}{4}\ell + 50
	\bigg] \theta_{n3}\nn\\
& &	-\,\sum_{\ell=1}^{n-2} \cfl\canlm\z2\left[
		3n^4 
		+ 6n^3 
		- (6\ell + 69)n^2 
		+ (18\ell + 72)n 
		- 12(\ell - 3)
	\right] \theta_{n3}\nn\\ 
& & +\,\sum_{\ell=2}^{n-2} \cfl\canlm\bigg[
		29n^4 
		- (27\ell + 17)n^3 
		+ \bigg(\,\frac{27}{4}\ell^2 + \frac{361}{4}\ell - 162\bigg)n^2
	\nn\\[-1mm]
& & \quad\quad
		-\bigg(\,\frac{189}{4}\ell^2 - \frac{269}{4}\ell - 164\bigg)n 
		+ 81\ell^2 - 293\ell - 8
	\bigg] \theta_{n4}
\Bigg\} \:\: .
\eea

\noindent
The corresponding results for $T_{\phi,\rm q}$ read
\bea
\lefteqn{ C^{\,(n,0)}_{\phi,\rm q} \ =\ -\,\frac{4^{n-2}}{6n!}\,\cf\,
\Bigg\{
	\cfnm\bigg(\,
		\frac{183}{2}n^2 
		+ \frac{3}{2}n 
		- 21
	\bigg)\ 
	+\ \cfnm\z2\left(
		48n^2 
		- 144n 
		+ 48
	\right)} \nn\\
& & -\,\cfnmm\ca\bigg(\,
		\frac{13}{2}n^2 
		- \frac{17}{2}n 
		+ 7
	\bigg)\theta_{n2}\ 
	-\ \cfnmm\ca\z2\left(
		54n^2 
		+ 6n 
		- 60
	\right)\theta_{n2} \nn\\[1mm]
& & +\,\cfnmm\b0\bigg(\,
		\frac{27}{4}n^3 
		+ \frac{13}{2}n^2 
		- \frac{55}{4}n 
		- \frac{1}{2}
	\,\bigg)\theta_{n2}
     +\,\sum_{\ell=0}^{n-3} \cfl\canlm\bigg[
		4n^2 
		+ 12n 
		+ \frac{27}{2}\ell^2 
\nn\\ & & \quad \mbox{}
                + \frac{53}{2}\ell - 8
	\bigg] \theta_{n3}
     -\,\sum_{\ell=0}^{n-3} \cfl\canlm\z2\left[
		6n^2 
		+ 42n 
		+ 12\ell - 36
	\right]\theta_{n3}\nn\\
& & +\,\sum_{\ell=0}^{n-3} \b0\cfl\canlmm\bigg[
		\bigg(\,\frac{9}{4}\ell + 2\bigg)n^2 
		+ \bigg(\,\frac{27}{4}\ell + 6\bigg)n 
		- \frac{9}{4}\ell^2 - \frac{25}{4}\ell - 4
	\bigg] \theta_{n3}\nn\\
& &	+\,\sum_{\ell=0}^{n-3} \bb02\cfl\canlmmm\bigg[\,
		\frac{3}{32}n^4 
		+ \frac{25}{48}n^3 
		- \frac{1}{32}(12\ell - 13)n^2 
		- \frac{1}{48}(54\ell + 121)n 
	\nn\\[-1mm]
& &	\quad\quad 
		+ \frac{1}{8}(\ell^2 + 3\ell + 2)
	\bigg] \theta_{n3}
\Bigg\} \:\: ,
\\[2mm]
\lefteqn{ C^{\,(n,1)}_{\phi,\rm q} \ =\ \frac{4^{n-2}}{6n!}\,\cf\,
\Bigg\{
	\cfnm\bigg(\,
		\frac{183}{2}n^3 
		- 9n^2 
		- \frac{171}{2}n 
		+ 3
	\bigg)\
	+\ \cfnm\z2\left(
		48n^3 
		- 192n^2 
		+ 216n 
		- 72
	\right)}\nn\\
& & -\,\cfnmm\ca\bigg(\,
		\frac{13}{2}n^3 
		- 34n^2 
		+ \frac{253}{2}n 
		- 65
	\bigg)\theta_{n2}\ 
	-\ \cfnmm\ca\z2\left(
		54n^3 
		- 60n^2 
		- 78n 
		+ 84
	\right)\theta_{n2}\nn\\[1mm]
& & +\,\cfnmm\b0\bigg(\,
		\frac{27}{4}n^4 
		+ \frac{31}{2}n^3 
		- \frac{343}{4}n^2 
		+ \frac{91}{2}n 
		- 32
	\bigg)\,\theta_{n2}\nn\\
& & +\,\sum_{\ell=0}^{n-3} \cfl\canlm\bigg[
		4n^3 
		+ \bigg(\,\frac{27}{2}\ell^2 + \frac{53}{2}\ell - 28\bigg)n
		+ \frac{27}{2}\ell^2 - \frac{43}{2}\ell - 32
	\bigg] \theta_{n3}\nn\\
& & -\,\sum_{\ell=0}^{n-3} \cfl\canlm\z2\bigg[
		6n^3 
		+ 24n^2 
		+ (12\ell-90)n 
		- 12\ell + 60
	\bigg] \theta_{n3}\nn\\
& & 	+\,\sum_{\ell=0}^{n-3} \b0\cfl\canlmm\bigg[
		\bigg(\,\frac{9}{4}\ell + 2\bigg)n^3 
		+ \bigg(\,\frac{45}{4}\ell - 12\bigg)n^2 
		- \bigg(\,\frac{9}{4}\ell^2 + \frac{7}{4}\ell + 26\bigg)n
	\nn\\ 
& & 	\quad\quad
		- 18\ell^2 - 22\ell - 4
	\bigg] \theta_{n3}\nn\\
& & +\,\sum_{\ell=0}^{n-3} \bb02\cfl\canlmmm\bigg[\,
		\frac{3}{32}n^5 + \frac{77}{96}n^4 
		- \frac{1}{96}(36\ell - 37)n^3 
		- \frac{1}{96}(216\ell + 533)n^2 
	\nn\\
& & \quad\quad
		+ \frac{1}{4}\bigg(\,\frac{1}{2}\ell^2 - 6\ell 
                   + \frac{1}{12}\,\bigg)n 
		+ \frac{1}{8}(23\ell^2 + 69\ell + 46)
	\bigg] \theta_{n3}
\Bigg\}
\eea
and
\bea
\lefteqn{ C^{\,(n,2)}_{\phi,\rm q} \ =\ -\,\frac{4^{n-2}}{6n!}\,\cf\,
\Bigg\{
	\cfnm\bigg(\,
		\frac{183}{4}n^4 
		- \frac{111}{2}n^3 
		- \frac{531}{4}n^2 
		+ \frac{201}{2}n 
		+ 42
	\bigg)}\nn\\
& & +\,\cfnm\z2\left(
		24n^4 
		- 144n^3 
		+ 312n^2 
		- 288n 
		+ 96
	\right)
    -\,\cfnmm\ca\bigg(\,
		\frac{13}{4}n^4 
		- 33n^3 
\nn\\ & & 
		+ \frac{633}{4}n^2 
		- \frac{463}{2}n 
		+ 42
	\bigg)\theta_{n2}
    -\,\cfnmm\ca\z2\left(
		27n^4 
		- 90n^3 
		+ 27n^2 
		+ 144n 
		- 108
	\right)\theta_{n2}\nn\\[1mm]
& & +\,\cfnmm\b0\bigg(\,
		\frac{27}{8}n^5 
		+ \frac{71}{8}n^4 
		- \frac{819}{8}n^3 
		+ \frac{1413}{8}n^2 
		- 92n 
		+ \frac{93}{2}
	\,\bigg)\theta_{n2}\nn\\
& & +\,\sum_{\ell=0}^{n-3} \cfl\canlm\bigg[
		2n^4 
		- 8n^3 
		+ \left(\frac{27}{4}\ell^2 + \frac{53}{4}\ell - 10\right)n^2 
	\nn\\
& & \quad\quad
		+ \bigg(\,\frac{27}{4}\ell^2 - \frac{139}{4}\ell\bigg)n 
		- 27\ell^2 + 3\ell + 72
	\bigg] \theta_{n3}\nn\\
& & -\,\sum_{\ell=0}^{n-3} \cfl\canlm\z2\left[
		3n^4 
		+ (6\ell - 63)n^2 
		- (18\ell - 144)n 
		+ 12\ell - 84 
	\right]\theta_{n3}\nn\\
& & +\,\sum_{\ell=0}^{n-3} \b0\cfl\canlmm\bigg[
		\bigg(\,\frac{9}{8}\ell + 1\bigg)n^4 
		+\bigg(\,\frac{27}{4}\ell - 16\bigg)n^3 
		-\bigg(\,\frac{9}{8}\ell^2 + \frac{35}{4}\ell - 7\bigg)n^2 
	\nn\\
& & \quad\quad
		-\bigg(\,\frac{135}{8}\ell^2 + \frac{241}{8}\ell - 36\bigg)n 
		+\frac{117}{4}\ell^2 + \frac{165}{4}\ell + 12
	\bigg] \theta_{n3}\nn\\
& & +\,\sum_{\ell=0}^{n-3} \bb02\,\cfl\canlmmm\bigg[\,
		\frac{3}{64}n^6 
		+ \frac{95}{192}n^5 
		- \frac{1}{64}(12\ell + 23)n^4 
		- \frac{1}{192}(288\ell + 907)n^3 
	\nn\\
& & \quad\quad
		+ \frac{1}{16}(\ell^2 + 79)n^2 
		+ \frac{1}{48}(135\ell^2 + 567\ell + 329)n 
		- \frac{1}{8}(35\ell^2 + 105\ell + 70) 
	\bigg] \theta_{n3}
\Bigg\} \:\: .\qquad
\eea

%
\section{Resummation of the unfactorized expressions for $F_L$}
\label{sec:resummation2}
\setcounter{equation}{0}
Up to one additional power of $\ep$ the moments of the unfactorized 
longitudinal structure functions $T_{\:\!L,\rm q}$ and $T_{\:\!L,\rm g}$ are 
built up from $D$-dimensional exponentials in the same way as $T_{\:\!2,\rm g}$ 
and~$T_{\phi,\rm q}$, 
\beq
\label{TLqgABC}
  T^{\,(n)}_{L,k}(N) \;=\;
  \frac{1}{N^{\,1+\delta_{k\rm g}}\:\ep^{\:\!2n-2}}\: \sum_{i=0}^{n-1}\,
  \left(
         A^{\,(n,i)}_{L,k} \:+\:
  \ep\,  B^{\,(n,i)}_{L,k} \:+\:
  \eps\, C^{\,(n,i)}_{L,k} \:+\: \ldots\,
  \right)\, \exp (\ep\, (n-i)\, \ln N) \; .
\eeq
In these cases also the $\ep^{-n}$ poles vanish at order $\as^{\,n}$ which 
compensates the absence of $\ep^{-2n+1}$ contributions. Consequently 
Eqs.~(\ref{A-KLN}), (\ref{B-KLN}) and (\ref{C-KLN}) are valid also for $\FL$.
As discussed in the introduction, our main objective for $\FL$ is the 
resummation of $\,C_{L,\rm g}$. As in the case of $\Ftwo\,$ the corresponding 
$D$-dimensional quark coefficient function, and thus 
$T_{L,\rm q}$, is also required for this.

Also here the LL coefficients $A^{\,(n,0)}_{L,k}$ are very simple and closely 
related, 
\beq
  A^{\,(n,0)}_{L,\rm q} \;=\;
    \frac{2^{\:\!2\:\!n}}{(n-1)!}\; C_F^{\,n} \; , \quad
  A^{\,(n,0)}_{L,\rm g} \;=\;
    \frac{2^{\:\!2\:\!n+1}}{(n-1)!}\; C_A^{\,n-1} \nf \; .
\eeq
The corresponding NLL contributions to Eq.~(\ref{TLqgABC}) are given by
\bea
B^{\,(n,0)}_{L,\rm q} &\!=\!& \frac{4^{n-2}}{(n-1)!}\:\cfnm
\left\{\,
        \b0(n - 1)(n - 2)\
        +\ 16\,\ca(n - 1)(1 - \z2)
        \right.\nn\\
& & \left.\quad\quad\quad
        -\ 4\,\cf(9n - 13)\
        +\ 32\,\cf\z2(n - 1)
\right\} \:\: ,
\\[3mm]
B^{\,(n,1)}_{L,\rm q} &\!=\!& -\,\frac{4^{n-2}}{(n-2)!}\:\cfnm
\left\{
        \b0(n^2 + 3n - 6)\
        +\ 16\,\ca(n - 1)(1 - \z2)
        \right.\nn\\
& & \left.\quad\quad\quad
        -\ 4\,\cf(9n - 16)\
        +\ 32\,\cf\z2(n - 1)
\right\}
\eea
and
\bea
B^{\,(n,0)}_{L,\rm g} &\!=\!& \frac{4^{n-2}}{(n-1)!}\:\nf\canmm
\left\{
        2\,\b0(n - 1)(n - 2)\theta_{n2}\
        -\ 8\,\cf(2n - 1)\theta_{n2}\
        +\ 64\,\ca n
\right\}\nn \\
& & -\, 2\,\frac{4^{n-1}}{(n-1)!}\:\nf\sum_{l=2}^{n-1}\cfl\canlm\,\theta_{n3}
\:\: ,
\\[4mm]
B\,^{(n,1)}_{L,\rm g} &\!=\!& -\,\frac{4^{n-2}}{(n-1)!}\:\nf\canmm
\left\{
        2\,\b0(n^2 - n + 2)(n - 1)\theta_{n2}\
        -\ 8\,\cf n(2n - 3)\theta_{n2}\
        \right.\nn \\[-1mm] & & \left.\quad\quad \mbox{}
        +\ 64\,\ca n(n - 1) \right\}
  \; +\; 2\,\frac{4^{n-1}}{(n-1)!}\:\nf\sum_{l=2}^{n-1}\cfl\canlm(n - 2)\,
 \theta_{n3} \:\: .
\eea
The NNLL terms for $T_{L,\rm q}$ read
\bea
\lefteqn{ C^{\,(n,0)}_{L,\rm q} \ =\ \frac{4^{n-2}}{12(n-1)!}\:\cfnmm\,
\bigg\{
	\bb02\bigg(\,
		\frac{3}{8}n^4 
		+ \frac{7}{12}n^3 
		- \frac{7}{8}n^2 
		- \frac{85}{12}n 
		+ 7
	\bigg)}\nn \\[1mm]
& & +\,\cas\left(
		96n^2
		- 288n
		+192
	\right)\left(\z2 - \z3\right)
	+\ \b0\ca\left(
		12n^3
		- 84n
		+ 72
	\right)(1 - \z2)\nn\\[2mm]
& & -\,\b0\cf\left[
		3(9 - 8\z2)n^3
		- 32n^2 
		- (293 - 168\z2)n
		+ 2(149 - 72\z2)
	\right]\nn\\[2mm]
& & +\,6\,\cfs\left[
		(17 + 48\z2 - 64\z3)n^2
		- (463 + 112\z2 - 544\z3)n
		+ 510 + 32\z2 - 480\z3
	\right]\nn\\[1mm]
& & -\,8\,\cf\ca\left[
		(4 + 45\z2 - 48\z3)n^2
		- (134 + 111\z2 - 276\z3)n 
		+ 130 + 66\z2 - 228\z3
	\right]
\bigg\} \:\: , \quad
\\[5mm]
\lefteqn{ C^{\,(n,1)}_{L,\rm q} \ =\ -\,\frac{4^{n-2}}{12(n-1)!}\:\cfnmm\,
\bigg\{
	\bb02\bigg(\,
		\frac{3}{8}n^5
		+ \frac{41}{24}n^4
		- \frac{203}{24}n^3
		- \frac{41}{24}n^2 
		+ \frac{325}{12}n
		- 19
	\bigg)}\nn\\[1mm]
& & +\,\cas\left(
		96n^3
		- 384n^2
		+ 480n
		- 192
	\right)(\z2 - \z3)\nn \\[2mm] 
& & +\,\b0\ca\left(
		12n^4
		+ 12n^3
		- 228n^2 
		+ 420n
		- 216
	\right)(1 - \z2)\nn\\[2mm]
& & -\,\b0\cf\left[
		3(9 - 8\z2)n^4
		- 2(7 + 12\z2)n^3
		- 19(29 - 24\z2)n^2
		+ 2(649 - 420\z2)n
\right. \nn \\[1.5mm] & & \left. \quad\quad \mbox{}
		- 8(95 - 54\z2) 
	\right]
   \;+\;6\,\cfs\left[
		(17 + 48\z2 - 64\z3)n^3
		- (534 + 112\z2 - 608\z3)n^2
	\right.\nn\\[2mm]
& & \quad\quad\left. \mbox{}
		+ (1227 + 80\z2 - 1120\z3)n 
		- 710 - 16\z2 + 576\z3
	\right]\nn \\[2mm]
& &	-\,8\,\cf\ca\left[
		(4 + 45\z2 - 48\z3)n^3
		- (152 + 144\z2 - 324\z3)n^2
	\right.\nn\\
& & \quad\quad\left.	
		+ (308 + 165\z2 - 540\z3)n
		- 160 - 66\z2 + 264\z3
	\right]
\bigg\}
\eea
and
\pagebreak
\bea
\lefteqn{ C^{\,(n,2)}_{L,\rm q} \ =\ \frac{4^{n-2}}{12(n-1)!}\:\cfnmm\,
\bigg\{ 
	\bb02\bigg(\,
		\frac{3}{16}n^6
		+ \frac{59}{48}n^5
		- \frac{151}{16}n^4
		+ \frac{281}{48}n^3
		+ \frac{183}{4}n^2
		- \frac{1039}{12}n
		+ 43
	\bigg)}\nn\\[0.5mm]
& &	+\,\cas\left(
		48n^4
		- 288n^3
		+ 624n^2
		- 576n 
		+ 192
	\right)(\z2 - \z3)\nn\\[1.5mm]
& &	+\,\b0\ca\left(
		6n^5
		+ 6n^4
		- 222n^3
		+ 714n^2
		- 864n
		+ 360
	\right)(1 - \z2)\nn\\
& & -\,\b0\cf\bigg(\,
		\frac{3}{2}(9 - 8\z2)n^5
		- \frac{1}{2}(23 + 24\z2)n^4
		- \frac{3}{2}(313 - 296\z2)n^3
		+ \frac{3}{2}(1289 - 952\z2)n^2
	\nn\\
& & \quad\quad	
		-\ 24(115 - 72\z2)n
		+ 2(647 - 360\z2)
	\bigg)\nn\\ 
& & +\,3\,\cfs\left[
		(17 + 48\z2 - 64\z3)n^4
		- 2(311 + 80\z2 - 368\z3)n^3
		+ (2675 + 144\z2 - 2432\z3)n^2
	\right.\nn\\[1.5mm]
& & \quad\quad\left.	
		- 2(1963 + 16\z2 - 1552\z3)n
		+ 64(29 - 21\z3)
	\right]\nn\\[1mm]
& &	-\,4\cf\ca\left[
		(4 + 45\z2 - 48\z3)n^4
		- 6(29 + 37\z2 - 70\z3)n^3
		+ 3(228 + 139\z2 - 408\z3)n^2
	\right.\nn\\
& & \quad\quad\left.	
		- 6(149 + 62\z2 - 242\z3)n
		+ 4(95 + 33\z2 - 150\z3)
	\right] 
\bigg\} \:\: .
\eea
The corresponding (and more complicated) coefficients for the gluon case are
\bea
\lefteqn{ C^{\,(n,0)}_{L,\rm g} \ =\ \frac{4^{n-2}}{(n-1)!}\:\nf\, 
\bigg\{
	\canm\bigg(\,
		\frac{4}{3}(77 - 15\z2)n^2 
		- 4(5 + \z2)n 
		+ \frac{8}{3}(17 - 3\z2)
	\bigg)}\nn \\
& & -\,8\cf\,\delta_{n2} 
	- 2(4\cf\b0 - 31\cfs + 16\cfs\z2)\,\delta_{n3}
	+ \canmm\b0\bigg(
		4n^3 
		- \frac{26}{3}n^2 
		- 2n 
		+ \frac{20}{3}
	\,\bigg)\theta_{n2}\nn\\[1mm]
& &	-\,\canmm\cf\left[
		16(5 - \z2)n^2 
		- 8(29 - 5\z2)n 
		+ 8(20 - 3\z2)
	\right]\theta_{n3}\nn\\
& & -\,\canmmm\dabcna\,\flg11\, 
		(32n^2 - 96n + 64)(11 + 2\z2 - 12\z3)\,
	\theta_{n3}\nn\\
& &	+\,\canmmm\bb02\bigg(\,
		\frac{1}{16}n^4 
		- \frac{29}{72}n^3 
		+ \frac{41}{48}n^2 
		- \frac{49}{72}n 
		+ \frac{1}{6}
	\,\bigg)\theta_{n3}\nn\\
& & +\,\cfnm\left[
		2(9 - 8\z2)n 
		- 16(2 - \z2)
	\right]\theta_{n4}
	-\ \cfnmm\b0\bigg(\,
		\frac{1}{2}n^2 
		+ \frac{1}{2}n 
		- 2
	\bigg)\theta_{n4}\nn\\[1mm]
& & +\,\canmmm\cfs\left[
		20n^2 
		- 8(7 + \z2)n 
		+ 2(17 + 4\z2)
	\right]\theta_{n4}\nn\\
& & -\,\canmmm\cf\b0\bigg(
		n^3 
		- \frac{7}{2}n^2 
		+ \frac{9}{2}n - 1
	\bigg)\theta_{n4}\
	-\ \sum_{\ell=2}^{n-3} \cfl\canlmm\b0\bigg(\,
		\frac{1}{2}n^2 
		- \frac{1}{2}n 
		+ \ell
	\bigg)\theta_{n4}\nn\\[-1mm]
& & +\,\sum_{\ell=3}^{n-2} \cfl\canlm\left[
		4(1-2\z2)n 
		+ 6\ell - 2(9-4\z2)
	\right]\theta_{n4}
\Bigg\} \:\: ,
\\[4mm]
\lefteqn{ C^{\,(n,1)}_{L,\rm g} \ =\ -\,\frac{4^{n-2}}{(n-1)!}\:\nf\, 
\bigg\{
	16\cf\z2\delta_{n2}\ 
	-\ 8(3\cf\b0 - 12\cfs + 4\cfs\z2)\delta_{n3}}\nn\\
& &	+\,\canm\bigg(\,
		\frac{4}{3}(77 - 15\z2)n^3 
		- 8(16 - 3\z2)n^2 
		+ \frac{4}{3}(61 - 21\z2)n 
		- 8(7 - 3\z2)
	\bigg)\nn\\
& & +\,\canmm\b0\bigg(
		4n^4 
		- \frac{14}{3}n^3 
		- 24n^2 
		+ \frac{182}{3}n 
		- 36
	\bigg)\theta_{n2}\nn\\[1mm]
& &	-\,\canmm\cf\left[
		16(5 - \z2)n^3 
		- 8(39 - 7\z2)n^2 
		+ 24(16 - 3\z2)n 
		- 32(5 - \z2)
	\right]\theta_{n3}\nn\\
& & -\,\canmmm\:\dabcna\:\flg11 
		(32n^3 - 128n^2 + 160n - 64)(11 + 2\z2 - 12\z3)
	\theta_{n3}\nn\\
& & +\,\canmmm\bb02\bigg(\,
		\frac{1}{16}n^5 
		- \frac{31}{144}n^4 
		- \frac{59}{144}n^3 
		+ \frac{319}{144}n^2 
		- \frac{179}{72}n 
		+ \frac{5}{6}
	\,\bigg)\theta_{n3}\nn\\
& &	+\,\cfnm\left[
		2(9 - 8\z2)n^2 
		- 2(37 - 24\z2)n 
		+ 4(19 - 8\z2)
	\right]\theta_{n4}\nn\\
& &	-\,\canmmm\cf\b0\bigg(
		n^4 
		- \frac{5}{2}n^3 
		+ \frac{1}{2}n^2 
		+ 4n 
		- 6
	\bigg)\theta_{n4}\nn\\[1.5mm]
& &	+\,\canmmm\cfs\left[
		20n^3 
		- 4(19 + 2\z2)n^2 
		+ 2(43 + 12\z2)n 
		- 2(17 + 8\z2)
	\right]\theta_{n4}\nn\\
& &	-\,\cfnmm\b0\bigg(\,
		\frac{1}{2}n^3 
		+ \frac{1}{2}n^2 
		- 6n 
		+ 8
	\bigg)\theta_{n4}\
	-\ \sum_{\ell=2}^{n-3} \cfl\canlmm\b0\bigg(\,
		\frac{1}{2}n^3 
		- \frac{1}{2}n^2 
		+ \ell n 
		- 4\ell
	\bigg)\theta_{n4}\nn\\[-1mm]
& &	+\,\sum_{\ell=3}^{n-2} \cfl\canlm\left[
		4(1 - 2\z2)n^2 
		+ (6\ell - 2(13 - 12\z2))n 
		- 18\ell + 2(21 - 8\z2)
	\right]\theta_{n4}
\bigg\} \qquad
\eea
and finally
\bea
\lefteqn{ C^{\,(n,2)}_{L,\rm g} \ =\ \frac{4^{n-2}}{(n-1)!}\:\nf\, 
\bigg\{
	\canmm\b0\bigg(
		2n^5 
		- \frac{7}{3}n^4 
		- \frac{92}{3}n^3 
		+ 111n^2 
		- \frac{436}{3}n 
		+ \frac{196}{3}
	\bigg)\theta_{n2}}\nn\\
& & -\,4(3\cf\b0 - 10\cfs)\delta_{n3}
	\;+\;\canm\bigg(\,
		\frac{2}{3}(77 - 15\z2)n^4 
		- \frac{4}{3}(127 - 27\z2)n^3 
	\nn\\
& & \quad\quad 
		+ 2(91 - 29\z2)n^2 
		- \frac{8}{3}(49 - 27\z2)n 
		+ \frac{40}{3}(5 - 3\z2)
	\bigg)\nn\\
& & -\,\canmm\cf\left[
		8(5 - \z2)n^4 
		- 4(59 - 11\z2)n^3 
		+ 4(125 - 24\z2)n^2 
	\right.\nn\\[1.5mm]
& & \quad\quad\left.	
		- 4(114 - 25\z2)n 
		+ 40(4 - \z2)
	\right]\theta_{n3}\nn\\[1mm]
& & -\,\canmmm\,\dabcna\:\flg11 
		(16n^4 - 96n^3 + 208n^2 - 192n + 64)(11 + 2\z2 - 12\z3)
	\theta_{n3}\nn\\
& &	+\,\canmmm\bb02\bigg(\,
		\frac{1}{32}n^6 
		- \frac{13}{288}n^5 
		- \frac{79}{96}n^4 
		+ \frac{1217}{288}n^3 
		- \frac{65}{8}n^2 
		+ \frac{473}{72}n 
		- \frac{11}{6}
	\,\bigg)\theta_{n3}\nn\\[1.5mm]
& & +\,\cfnm\left[
		(9-8\z2)n^3 
		- (67-48\z2)n^2 
		+ 4(41-22\z2)n 
		- 12(11-4\z2)
	\right]\theta_{n4}\nn\\[1.5mm]
& & +\,\canmmm\cfs\bigg[
		10n^4 
		- 2(29+2\z2)n^3 
		+ 3(39+8\z2)n^2 
		- 11(9+4\z2)n 
		+ 8(5+3\z2)
	\bigg]\theta_{n4}\nn\\
& &	-\,\canmmm\cf\b0\bigg(
		\frac{1}{2}n^5 
		- \frac{5}{4}n^4 
		- 3n^3 
		+ \frac{51}{4}n^2 
		- 19n 
		+ 15
	\bigg)\theta_{n4}\nn\\[1.5mm]
& & -\,\cfnmm\b0\bigg(
		\frac{1}{4}n^4 
		- \frac{27}{4}n^2 
		+ \frac{39}{2}n 
		- 18
	\bigg)\theta_{n4}\nn\\
& &	-\,\sum_{\ell=2}^{n-3} \cfl\canlmm\b0\bigg(\,
		\frac{1}{4}n^4 
		- \frac{1}{2}n^3 
		+ \frac{1}{4}(2\ell - 5)n^2 
		- \frac{3}{2}(3\ell - 1)n 
		+ 9\ell
	\bigg)\theta_{n4}\nn\\
& &	+\,\sum_{\ell=3}^{n-2} \cfl\canlm\left[
		2(1-2\z2)n^3 
		+ (3\ell-19+24\z2)n^2 
		- (21\ell-63+44\z2)n 
	\right.\nn\\[-2mm]
& & \quad\quad\left.	
		+ 36\ell - 24(3-\z2)
	\right]\theta_{n4}
\Bigg\} \:\: .
\eea

%
\setcounter{equation}{0}
\section{NNLL resummation of the off-diagonal splitting functions}
\label{sec:NNLLsplit}
Together with the mass-factorization relations of Section 2, the results of 
the previous two sections facilitate the iterative determination of the 
respective coefficients $D_{\:\! ik}^{\,(n,\ell \,\leq\, 2)}$ and 
$D_{a,\, k}^{\,(n,\ell \,\leq\, 2)}$ in Eqs.~(\ref{Pij-xto1}), 
Eqs.~(\ref{Cak-xto1}) and Eqs.~(\ref{CLk-xto1}) to any order $\as^{\,n}$. 
In this section we present the resulting expressions for the splitting 
functions in Mellin-$N$ space, from which the $x$-space coefficients 
$D_{\:\!\rm qg}^{\,(n,\ell \,\leq\, 2)}$ and 
$\:\!D_{\rm gq}^{\,(n,\ell \,\leq\, 2)}$ can readily be obtained by inverting
the second relation in (\ref{LogsTrf}). The corresponding results for the
coefficient functions are discussed in the next section. 

The LL and NLL contributions to $P_{\:\!\rm qg}$ and $P_{\:\!\rm gq}$ can be
expressed in a closed form in terms of a new class of functions with Taylor
expansions in terms of Bernoulli numbers. Extending the definitions of
Ref.~\cite{AV2010} to $\,k = 2,\,3,\,\ldots\,$, these functions are given by
\beq
\label{Bkdef}
  \B{k}(x) \;=\;
  \sum_{n\,=\,0}^\infty \;\frac{B_n}{n!(n+k)!}\; x^{\,n} \, ,\quad
  \B{-k}(x) \;=\;
  \sum_{n\,=\,k}^\infty \;\frac{B_n}{n!(n-k)!}\; x^{\,n} \:\: .
\eeq
$B_n$ are the Bernoulli numbers in the standard normalization of 
Ref.~\cite{AbrSteg}: $B_{2n+1} =\, 0$ for $n \geq 1$ and
\beq
\label{Bnumbers}
  B_0 \:=\: 1 \:, \;\;
  B_1 \:=\: -{1 \over 2} \:, \;\;
  B_2 \:=\: {1 \over 6} \:, \;\;
  B_4 \:=\: -{1 \over 30} \:, \;\;
  B_6 \:=\: {1 \over 42} \:, \;\; \ldots , \;\;
  B_{12} \:=\: -{691 \over 2730} \:, \;\;\ldots
\;\; .
\eeq
The functions $\B{\pm k}$ for $k>0$ are related to $\B{0}$, which can also be
written as
\beq
\label{B0zeta}
  \B{0}(x) \;= \; 1 \,-\: {x \over 2}
  \; - \;\sum_{n=1}^\infty \,\frac{(-1)^n}{[(2n)!]^2} \; |B_{2n}|\, x^{\,2n}
  \;= \; 1 \:-\: {x \over 2} \:-\: 2 \sum_{n=1}^\infty \:
  {(-1)^n \over (2n)!} \,\zeta_{2n}^{} \Big( {x \over 2\pi} \Big)^{\!2n}
  \; , 
\eeq
 by
\beq
\label{Bkdiff}
  \frac{d^{\,k}}{dx^{\,k}}\; (x^{\,k} \B{k}) \;=\; \B0
\:\: , \quad
  \frac{d^{\,k}}{dx^{\,k}}\; \B0 \;=\; {1 \over x^{\,k}}\; \B{-k}
\:\: .
\eeq
Due to $\,\zeta_{2n} \ra 1$ for $\,n \ra \infty\,$ the series (\ref{Bkdef})
converge absolutely for all values on $x$.

\begin{figure}[p]
\centerline{\epsfig{file=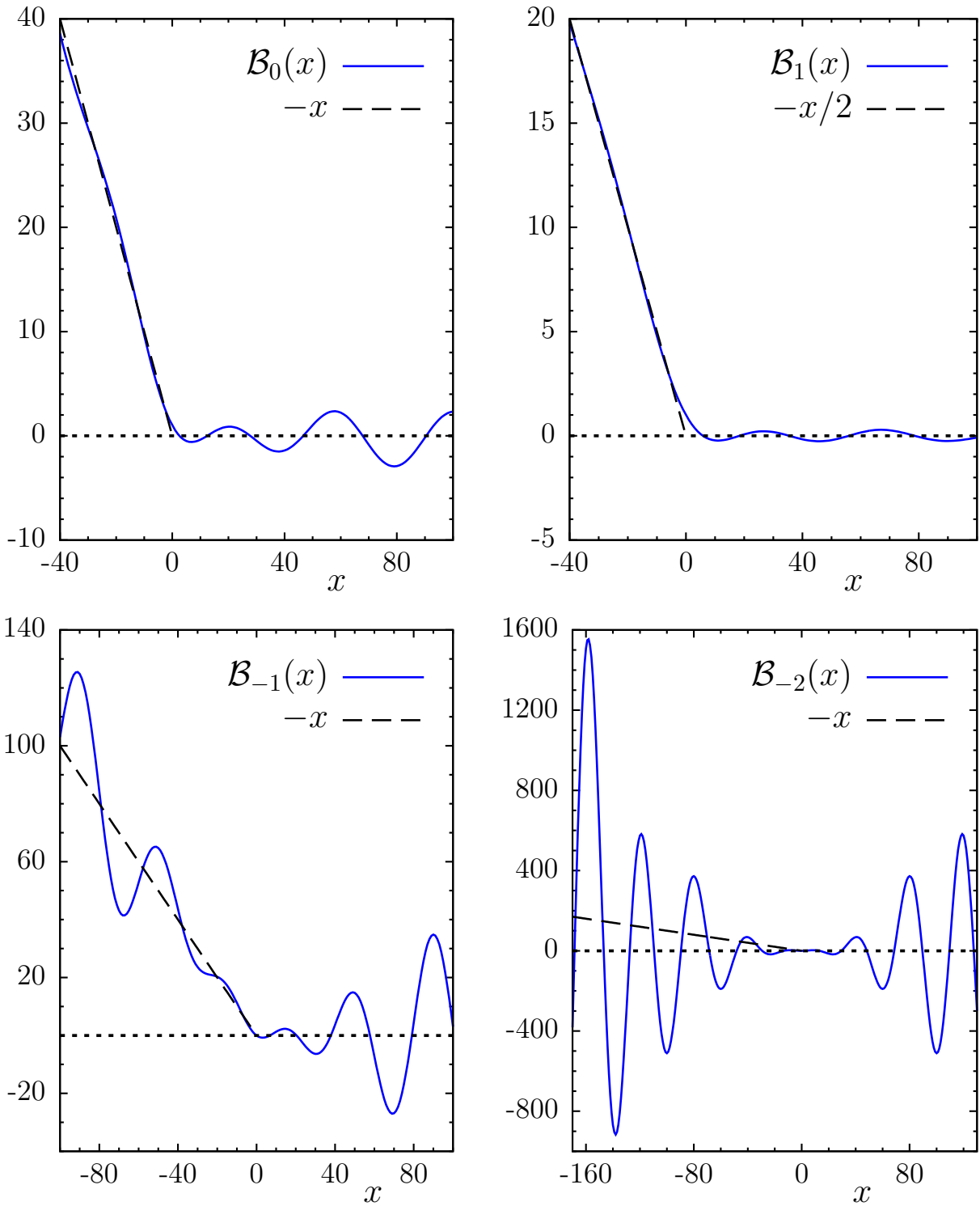,width=16cm,angle=0}}
\caption{The functions ${\cal B}_{k}(x)$, $k = 0,\, \pm 1,\, 2$, evaluated 
using their defining Taylor expansions~(\ref{Bkdef}).
\label{fig:Bfunc}}
\end{figure}

The numerical behaviour of the functions $\B{k}(x)$ with $k=0,\pm 1,-2$ which 
enter our results below is illustrated in Fig.~\ref{fig:Bfunc}. Similar to 
$\B0(x)$ -- where this oscillation has been shown to continue, albeit in a much
more irregular fashion, to much larger values of $x$ \cite{DBpriv} --
$\B{k\geq0}(x)$ oscillates around $y=0$ for positive $x$ and $y=-x/(k+1)!$ for 
negative $x$. On the other hand, $\B{k<0}(x)$ also oscillates around $y=0$ for 
positive $x$ but around $y=-x$ for negative $x$. As can be seen from the 
figures, the amplitude of these oscillations increases very rapidly with 
decreasing $k$.

The resummed gluon-quark splitting function at large $N$ can now be written as
\bea
\label{eq:pqgNL} 
 N\:\!P_{\rm qg}(N,\as) &\!=\!& 2\:\!\ar\,\nf\,\B0(\art) 
 \nn \\[0.5mm] & & \mbox{\hspn} 
 +\,\ar^{\,2} \ln\, \Ntil\: \nf 
    \Big[\left(12\cf-2\b0\right)\:\frac{1}{\art}\:\B{-1}(\art)
  \,+\: \frac{\b0}{\art}\:\B{-2}(\art) 
  \,+\, \left(6\cf-\b0\right) \B1(\art) \Big]
 \nn \\[0.5mm] & & \mbox{\hspn}
 +\; \mbox{ NNLL contributions } \;+\; \ldots \:\: . 
\eea
 
\pagebreak
\noindent
Here and below we use (recall $\,\ln\, \Ntil = \ln N + \GE$)
\beq
\label{atilde}
  \art \;\equiv\; 4\:\!\ar\,(C_A-C_F)\,\ln^2 \Ntil \; .
\eeq
The corresponding result for the quark-gluon splitting reads
\bea
\label{eq:pgqNL}
 N\:\!P_{\rm gq}(N,\as) &\!=\!& 2\:\!\ar\,\cf\,\B0(-\art) 
 \nn \\[0.5mm] & & \mbox{\hspn}
 +\,\ar^{\,2} \ln\, \Ntil\: \cf
    \Big[\left(12\:\!\cf-6\:\!\b0\right)\:\frac{1}{\art}\:\B{-1}(-\art)
  \,-\: \frac{\b0}{\art}\:\B{-2}(-\art) 
 \nn \\[-0.5mm] & & \mbox{\hspp}
 +\, \left(14\:\!\cf-8\:\!\ca-\b0\right)\:\B1(-\art)\Big]
 \nn \\[0.5mm] & & \mbox{\hspn}
 +\; \mbox{ NNLL contributions } \;+\; \ldots \:\: . 
\eea
The first lines of Eqs.~(\ref{eq:pqgNL}) and (\ref{eq:pgqNL}) are the 
respective LL results, derived in Ref.~\cite{AV2010} from relations equivalent
to Eq.~(\ref{ToffLL}).
Unlike for the NLL corrections, we have not been able to find closed relations
for all colour factors contributing to the NNLL terms. We therefore provide
these results in the form of tables to order $\as^{\:\!18}$ which can be
found in Appendix A. 

Our results for the fourth-order (N$^3$LO) splitting functions agree with the 
predictions of Ref.\ \cite{SMVV1} derived from the conjectured 
single-logarithmic large-$x$ enhancement of the physical evolution kernels 
for the system $(\Ftwo,\,F_\phi)$ of flavour-singlet structure functions.
Furthermore they show the expected extension of the colour-factor pattern
seen in those results to all orders in $\as$: the LL terms for both
$P_{\rm qg}^{\,(n)}$ and $P_{\rm gq}^{\,(n)}$ [recall Eq.~(\ref{P-exp})] 
are proportional to $(\cf\!-\ca)^n$, the NLL terms include at most two 
colour factors other that $(\cf\!-\ca)$, and the NNLL terms of Appendix A
involve $(\cf\!-\ca)^{n-2}$ or higher powers of $(\cf\!-\ca)$. It appears that
generally all double-logarithmic contributions, $\ln^{\,k} N$ with 
$n+1\leq k \leq 2\:\!n$, vanish for $\cf=\ca$ which is part of the 
colour-factor choice leading to an $\:\!{\cal N}=1$ supersymmetric theory.

It is clear already from the discussion of the fourth-order results in 
Ref.~\cite{SMVV1} (Figs.~12 and 13) that the highest three logarithms of
$P_{\rm qg}^{\,(n>2)}$ and $P_{\rm gq}^{\,(n>2)}$ can provide no more than
a very rough indication of the size of the corrections beyond NNLO
(note that the expansion in Ref.~\cite{SMVV1} is in terms of $\ln N$
instead of $\ln\, \Ntil$). 
Nevertheless it is useful to take a brief look at the numerical size and
convergence of the resummation corrections to these quantities. This is done
in Figs.~2 and 3 at the reference point
\beq
\label{as-ref}
  \as(\Qs) \;=\; 0.2 \:\: , \quad  \nf \;=\; 4 
\eeq
used before, e.g., in Refs.~\cite{MVV3,MVV4,MVV5,MVV6,MVV10}. Depending on the
precise value of $\as$ at the $Z$-boson mass, this choice corresponds to a 
scale $\Qs \approx 25 \ldots 50 \mbox{ GeV}^2$ typical for measurements of
DIS \cite{PDG2010}.

The figures show that the resummation corrections to the NNLO splitting 
functions are dominated by the third (NNL) logarithms which completely 
overwhelm the LL and NLL contributions (except, of course, for huge but 
practically irrelevant values of $N$). 
The relative corrections to the already small large-$N$ off-diagonal splitting 
functions are rather small, amounting to less than 2\% and about 3\% at 
$N = 20$ for for $P_{\rm qg}$ and $P_{\rm gq}$ respectively.
At least at the present level of logarithmic accuracy contributions beyond 
order $\as^{\,4}$ (N$^3$LO) are negligible.

\begin{figure}[p]
\vspace*{-1mm}
\centerline{\epsfig{file=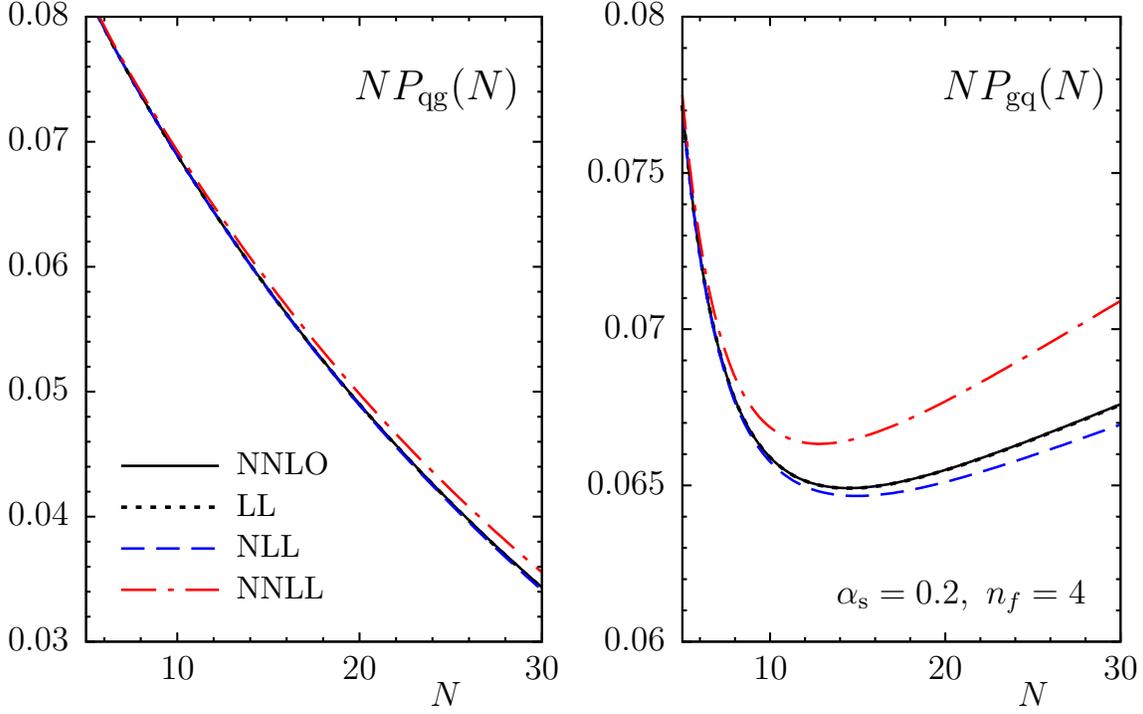,width=15cm,angle=0}}
\caption{The Mellin-$N$ splitting function $P_{\rm qg}$ and $P_{\rm gq}$,
multiplied by $N$ for display purposes. Shown are the LL, NLL and NNLL
large-$N$ resummation corrections to the complete NNLO results \cite{MVV4}.
\label{fig:splitt}
}
\end{figure}
 
\begin{figure}[p]
\vspace*{-1mm}
\centerline{\epsfig{file=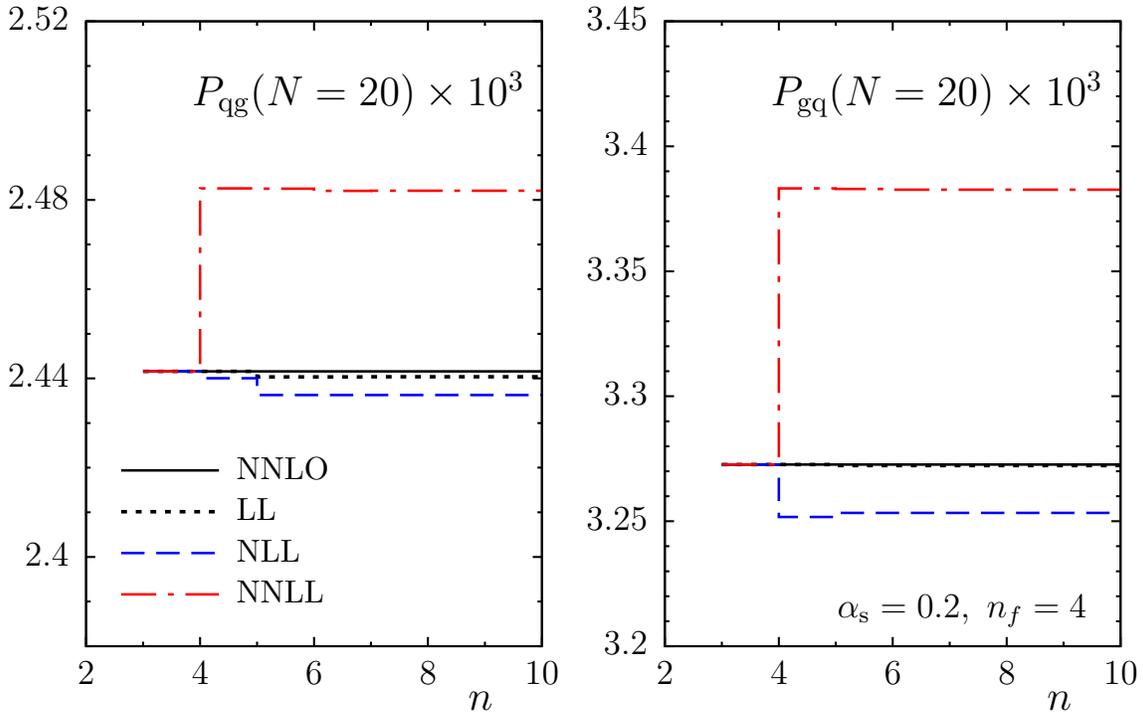,width=15cm,angle=0}}
\caption{The contributions of the various orders in $\as$ to the resummation 
corrections at $N = 20$. The LL, NLL and NNLL terms of order $\as^{\,n}$ are 
added at the corresponding values of the abscissa. 
\label{fig:splitconv}
}
\vspace*{-2mm}
\end{figure}
 
%
\setcounter{equation}{0}
\section{NNLL resummation of the coefficient functions}
\label{sec:NNLLcoeff}
The resummed results for the coefficient functions $C_{2,\rm g}$ and 
$C_{\phi,\rm q}$ are more complicated than those for the splitting functions
already at the leading-log level. Hence it is not surprising that we are not
able to give closed NNLL expressions for these quantities either at this point.
Postponing the NNLL contributions to Appendix B, the results are
\bea
 N\:\!C_{2,\rm g}(N,\as) &\!=\! & 
  \frac{1}{2\ln \,\Ntil^{}}\: \frac{\nf}{\ca-\cf}
  \left[\exp(2\ar\cf\ln^2\Ntil^{})\B0(\art)-\exp(2\ar\ca\ln^2\Ntil)\right]
\nn \\[1mm] & & \mbox{\hspn} -\: 
  \frac{1}{8\ln^2\Ntil}\: \frac{\nf(3\cf-\b0)}{(\ca-\cf)^2} 
  \left[\exp(2\ar\cf\ln^2\Ntil)\B0(\art)-\exp(2\ar\ca\ln^2\Ntil)\right]
\nn \\[1mm] & & \mbox{\hspn} -\: 
  \frac{\ar}{4}\: \frac{\nf}{\ca-\cf}\: 
  \exp\,(2\ar\ca\ln^2\Ntil)\left(8\ca+4\cf-\b0\right) 
\nn \\[1mm] & & \mbox{\hspn} -\:
  \frac{\ar}{4}\:\frac{\nf}{\ca-\cf}\: 
  \exp\,(2\ar\cf\ln^2\Ntil)\Big[-6\cf\B0(\art)-(6\cf-\b0)\B1(\art) 
\nn \\[-1mm] & & \mbox{\hspp} -\:
  (12\cf-4\b0)\:\frac{1}{\art}\:\B{-1}(\art)-\:\frac{\b0}{\art}\:\B{-2}(\art)
  \Big] 
\nn \\[-0.5mm] & & \mbox{\hspn} -
  \frac{\ar^2}{3}\,\b0\ln^2 \Ntil\,\frac{\nf}{\ca-\cf}
  \left[\ca\exp(2\ar\ca\ln^2\Ntil) - \cf\exp(2\ar\cf\ln^2\Ntil)\B0(\art)\right]
\nn \\[2mm] & & \mbox{\hspn} +\:
  \mbox{ NNLL contributions } \;+\; \ldots
\label{eq:c2gNL}
\eea
and
\bea
 N\:\!C_{\phi,\rm q}(N,\as) &\!=\!& 
  \frac{1}{2\ln \,\Ntil}\:\frac{\cf}{\cf-\ca}
  \left[\exp(2\ar\ca\ln^2\Ntil)\B0(-\art)-\exp(2\ar\cf\ln^2\Ntil)\right]
\nn \\[1mm] & & \mbox{\hspn} +\:
  \frac{1}{8\ln^2\Ntil}\:\frac{\cf(3\cf-\b0)}{(\cf-\ca)^2}
  \left[\exp(2\ar\ca\ln^2\Ntil)\B0(-\art)-\exp(2\ar\cf\ln^2\Ntil)\right]
\nn \\[1mm] & & \mbox{\hspn} +\:
  \frac{\ar}{4}\:\frac{\cf}{\cf-\ca}\:
  \exp\,(2\ar\cf\ln^2\Ntil)\left(12\ca-18\cf-\b0\right) 
\nn \\[1mm] & & \mbox{\hspn} +\:
  \frac{\ar}{4}\:\frac{\cf}{\cf-\ca}\,
  \exp\,(2\ar\ca\ln^2\Ntil)\Big[2\b0\B0(-\art)-(\b0-6\cf+8\caf)\B1(-\art) 
\nn \\[-1mm] & & \mbox{\hspp} -
  (4\b0-12\cf)\:\frac{1}{\art}\:B{-1}(-\art)-\:\frac{\b0}{\art}\:\B{-2}(-\art)
  \Big] 
\nn \\[-0.5mm] & & \mbox{\hspn} +\:
  \frac{\ar^2}{3}\:\b0\ln^2 \Ntil\:\frac{\cf}{\cf-\ca}
  \left[\ca\exp(2\ar\ca\ln^2\Ntil)\B0(-\art)-\cf\exp(2\ar\cf\ln^2\Ntil)\right]
\nn \\[2mm] & & \mbox{\hspn} +\:
  \mbox{ NNLL contributions } \;+\; \ldots \:\: .
\label{eq:cphiqNL}
\eea
Here the first lines are the respective LL results of Ref.~\cite{AV2010}. 
As in these terms, the \mbox{$(\ca-\cf)$} denominators
are generally cancelled by corresponding numerator factors as can be seen by
expanding all functions in powers of $\as$. Unlike for the splitting functions 
the double-logarithmic contributions to the coefficient functions do not vanish
for $\cf=\ca$. However, they can be expressed in terms of exponentials in this
case since all ${\cal B}$-functions have the argument (\ref{atilde}).

Our results for the quark coefficient function for $\FL$ completely agree with 
those of Ref.~\cite{MV3}. The complexity of the N$^n$LL resummed expression for 
the gluon coefficient function $C_{L,\rm g}$ is similar to that of 
$C_{\:\!2,\rm g}$ at N$^{n\!-\!1}$LL level. Hence it can be written down in a 
closed form at NNLL accuracy,
\bea
 N^{\,2}\:\!C_{L,\rm g}(N,\as) &\!=\!& 
 8\:\!\ar\,\nf\,\exp\,(2\:\!\ar\ca\ln^2 \Ntil) 
 \;+\; 4\:\!\ar\cf\, N\:\! C^{\,\rm LL}_{2,\rm g}(N,\as) 
\nn \\ & & \mbox{\hspn} +\: 
 16\:\!\ar^2 \ln\,\Ntil\, \nf \, \exp(2\ar\ca\ln^2 \Ntil) 
 \bigg[(4\:\!\ca-\cf) + \frac{1}{3}\:\ar\ln^2 \Ntil\, \ca\b0 \bigg]
\nn \\[0.5mm]\ & & \mbox{\hspn} +\:
 4\:\!\ar^2\ln \,\Ntil\, \cf \left[\, \b0+4\:\!\ca(1-\z2)-4\:\!\cf(3-2\,\z2) 
 \right] \,N\:\! C^{\,\rm LL}_{2,\rm g}(N,\as) 
\nn \\ & & \mbox{\hspn} +\:
 4\:\!\ar\cf\, N\:\! C^{\,\rm NLL}_{2,\rm g}(N,\as) \:+\:
 8\:\!\ar^2\nf \, \exp(2\ar\ca\ln^2 \Ntil) \bigg[ \cf(1-2\z2) 
\nn \\[-0.5mm] & & \mbox{\hspp} 
 +\frac{1}{3}\:\ar\ln^2 \Ntil\, \bigg(\b0(22\:\!\ca-3\cf)+2\:\!\cas(79-18\,\z2)
 +30\,\cfs
\nn \\[-1mm] & & \mbox{\hspp\hspp}
 -24\:\!\ca\cf(5-\z2) 
 -48\:\dabcna\:\flg11(11+2\,\z2-12\,\z3)\bigg) 
\nn \\[-1mm] & & \mbox{\hspp}
 +\frac{1}{3}\:\ar^2\ln^4 \Ntil\: \b0\ca(\b0+16\:\!\ca-4\cf)
 +\frac{2}{9}\:\ar^3\ln^6 \Ntil\: \bb02\cas\bigg] \;+\; \ldots \:\: . \quad
\label{eq:cLgNL}
\eea
Here the first term is the LL result also obtained already in Ref.~\cite{MV3}.
The next term, where $C^{\,\rm LL}_{2,\rm g}$ stands for the first line of
Eq.~(\ref{eq:c2gNL}), and the second line form the NLL contribution.
The remainder of Eq.~(\ref{eq:cLgNL}) collects the NNL logarithms, where
$C^{\,\rm NLL}_{2,\rm g}$ represents the second to sixth line of 
Eq.~(\ref{eq:c2gNL}). Furthermore this is the only NNLL contribution including
the charge factor 
\beq
\label{flg11}
  fl_{11}^{\:\rm g}  \:=\: \langle e \rangle ^2 / \langle e^{\,2} \rangle 
  \quad \mbox{ with } \quad
  \langle e^{\,k} \rangle \:=\: \nf^{\!\!-1}\: \textstyle 
  \sum_{\,i=1}^{\,\nf}\: e_i^{\,k}
\eeq
(where $e_i$ is the charge of the $i$-th effectively massless flavour in units
of the proton charge) arising from diagrams with the colour factor 
$d^{\:\!abc}d_{abc}/n_a = 5/48\;\nfs$ in QCD where the two (neutral) gauge 
bosons couple to different quark loops of the gauge-bosons gluon forward 
amplitude \cite{Larin:1997wd,MVV5,MVV6}.

The NLL and NNLL contributions to Eq.~(\ref{eq:cLgNL}) for $\cf=0$ and 
$fl_{11}^{\:\rm g}=0$ agree with the previous all-order result \cite{MVV3} 
in this gluonic `non-singlet' limit to which $C^{\,\rm (N)LL}_{2,\rm g}$ and 
hence the special functions (\ref{Bkdef}) do not contribute. The complete
fourth-order coefficient function (for $W$-exchange, i.e., without the 
$fl_{11}^{\:\rm g}=0\,$ part) has been predicted in Ref.~\cite{LL2010} from
the physical evolution kernel for the system $(\Ftwo,\,\FL)$ of 
flavour-singlet structure functions together with the four-loop 
splitting-function results of Ref.~\cite{SMVV1}. 
The present results are in full agreement also with that prediction.

Eqs.~(\ref{eq:pqgNL}) and (\ref{eq:pgqNL}) for the resummed splitting 
functions together with their counterparts (\ref{eq:c2gNL})~--~%
(\ref{eq:cLgNL}) for the coefficient functions and the corresponding 
simpler results for $C_{2,\rm q}$, $C_{\phi,\rm g}$ and $C_{L,\rm q}$
facilitate an assumption-free NNLL calculation of the physical kernels for 
$(\Ftwo,F_\phi)$ and $(\Ftwo,\FL)$ to any order in $\as$. It turns out 
that their highest double logarithms, as far as they can be determined from 
available fixed-order results now, do indeed vanish. Hence the conjectures of 
Refs.~\cite{MV3,MV5,SMVV1,LL2010} are proven by our present calculations for
the leading $N^{\,-1}$ large-$N$ contributions.

The numerical size and $\as$-convergence of the LL, NLL and NNLL resummation
corrections to the respective third-order results are illustrated in Fig.~4
for $C_{2,\rm g}$ and Figs.~5 and 6 for the coefficient functions of $\FL$.
For brevity we do not show the corresponding results for $C_{\phi,\rm q}$
which is of theoretical but not phenomenological interest. 
The corrections are dominated by the NNLL contributions, suggesting that 
they underestimate the impact of at least very high orders.
Already the known terms, e.g., at $N=20$ for the reference point 
(\ref{as-ref}), are sizeable for $C_{L,\rm q}$ with about 15\%, large for 
$C_{2,\rm g}$ (about 35\%) and huge for $C_{L,\rm g}$ with about 100\%.
The gluonic quantities receive significant contributions from the fourth to 
sixth order in $\as$. With values of 10.2 and 49.5 the forth-order 
coefficients are comparable to the corresponding (averaged) Pad\'e estimates,
\bea
\label{PadeC2}
  -\,C_{2,\rm g}(N=20) &\!=\!&
   \,0.127\,\as \:+\: 0.642\,\as^{\,2} \:+\: 2.76\,\as^{\,3}
  \:+\: 12_{\,\rm Pad\acute{e}\,}\,\as^{\,4} \:+\: \ldots \; , 
\\[0.5mm]
\label{PadeCL}
  N\:\!C_{L,\rm g}(N=20) &\!=\!&
  \,0.110\,\as \:+\: 1.240\,\as^{\,2} \:+\: 9.51\,\as^{\,3} 
  \:+\: 65_{\,\rm Pad\acute{e}\,}\,\as^{\,4} \:+\: \ldots \; .
\eea

\begin{figure}[t]
\centerline{\epsfig{file=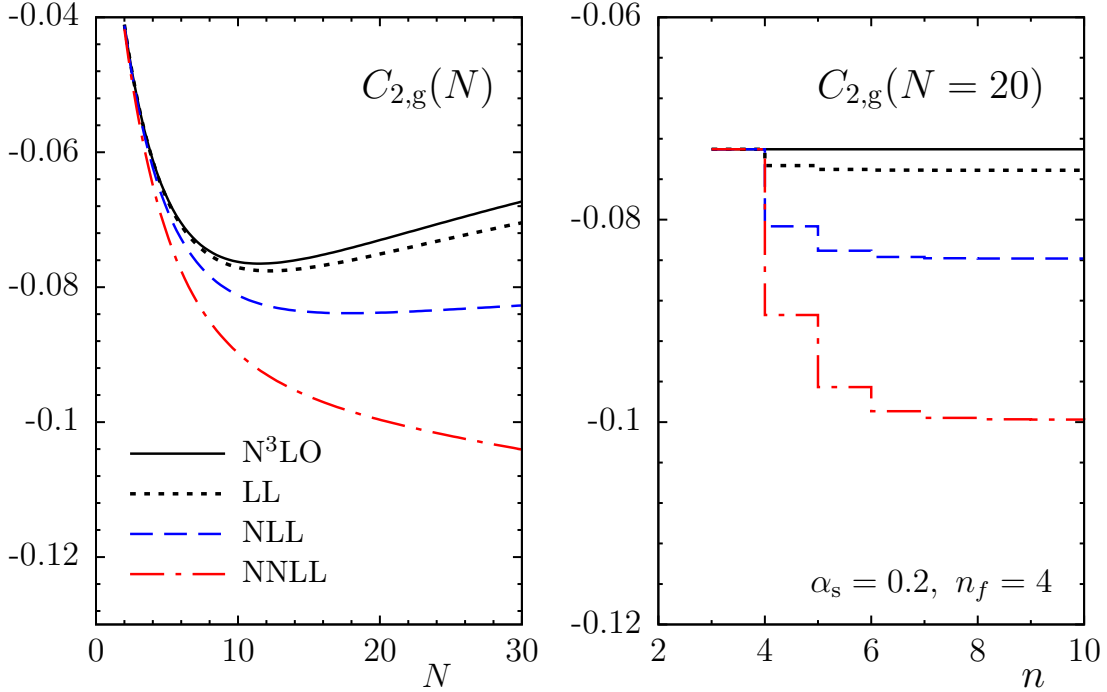,width=14.5cm,angle=0}}
\vspace*{-1mm}
\caption{Left: the $N$-dependence of the third-order (N$^3$LO) and resummed 
gluon coefficient function for $\Ftwo$ at the reference point (\ref{as-ref}). 
Right: the contributions of the fourth to the tenth orders in $\as$, added at 
the corresponding values of the abscissa, to those results at $N=20$.
}
\vspace*{-1mm}
\end{figure}

\begin{figure}[p]
\centerline{\epsfig{file=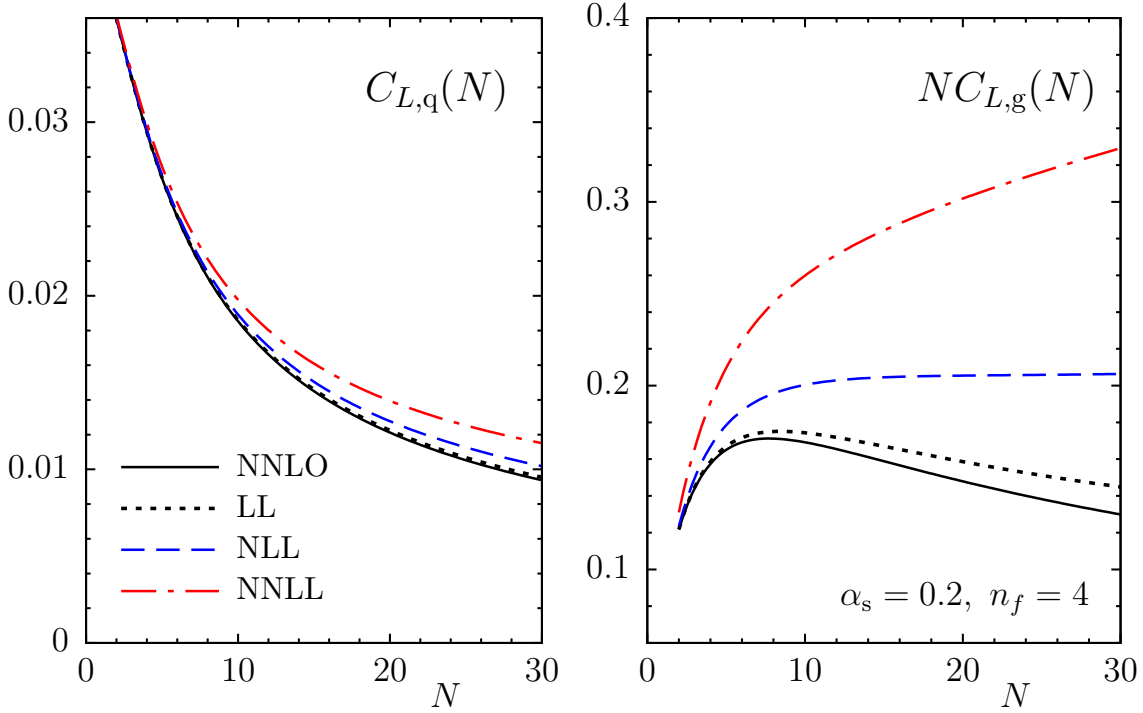,width=15cm,angle=0}}
\vspace*{-1mm}
\caption{The Mellin-$N$ space quark (left) and gluon (right, multiplied by 
$N$) coefficient functions for $\FL$ at the reference point (\ref{as-ref}). 
Shown are the cumulative LL, NLL and NNLL large-$N$ resummation corrections to 
the third-order (NNLO) results.
}
\end{figure}
\begin{figure}[p]
\centerline{\epsfig{file=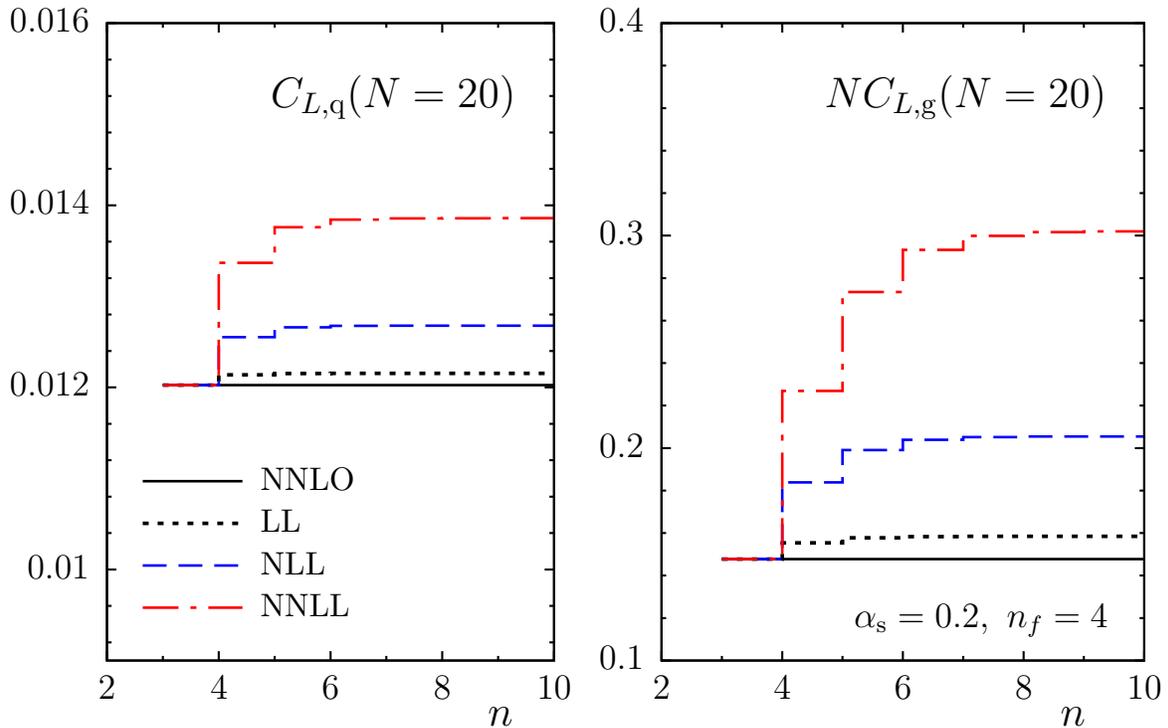,width=15.3cm,angle=0}$\;\;$}
\vspace*{-1mm}
\caption{As Fig.~3, but for the coefficient functions $C_{L,\rm q}$ and
$C_{L,\rm g}$ shown in Fig.~5 above.
}
\end{figure}

\vspace*{-2mm}
It is obvious from Eqs.~(\ref{Treal}) -- (\ref{Ttotal}) that our method can 
be applied as well to non-leading terms in the expansion in powers of
$\x1$. This is especially interesting for the subleading $\x1^0$ or $1/N$ 
contributions to diagonal coefficient function such as $C_{2,\rm q}$ and
$C_{\phi,\rm g}$. Due to the stable form (\ref{Pii-xto1}) of the diagonal
splitting functions, the corresponding N$^3$LO corrections to the non-singlet 
structure functions are known except for the single- and non-logarithmic
contributions. Consequently, as discussed below Eq.~(\ref{Ttotal}), we are
able to resum highest four $N^{\,-1}$ logarithms.

The resummation of the $N^{\,-1}$ non-singlet coefficient functions for 
$F_{\:\!1,2,3}$ has been inferred in Ref.~\cite{MV5} from the behaviour of the 
physical evolution kernels, obtaining complete NNLL results and the N$^3$LL 
corrections up to one undetermined number called $\xi_{\,\rm DIS_4}$. 
Hence it was not necessary to perform a very cumbersome all-order N$^3$LL 
calculation analogous to Section 3. Instead we have verified those previous 
results to order $\as^{\,7}$ and determined the hitherto missing parameter,
\beq
   \xi_{\,\rm DIS_4} \;=\; 100/3
\quad \mbox{ in Eqs.~(5.24) -- (5.26) of Ref.~\cite{MV5} }.
\eeq
 
%
\setcounter{equation}{0}
\section{Summary and outlook}
\label{sec:summary}
Unlike the case of the dominant $\x1^{-1}_+/\,N^{\,0}$ large-$x /\,$large-$N$ 
terms, only a small amount of (published) research has been devoted until 
recently to the all-order structure of $1/N$-suppressed threshold contributions 
to the coefficient functions for the structure functions in deep-inelastic 
scattering (DIS) and crossing-related (semi-)$\,$inclusive quantities in 
perturbative QCD. 
See Ref.~\cite{KLM96} for a well-known leading-log conjecture, and 
Refs.~\cite{Akhoury:1998gs,Akhoury:2003fw} for early studies of the 
longitudinal structure function $\FL$. The study of these contributions is 
however not only theoretically interesting but also phenomenologically 
relevant, e.g., for assessing the kinematic region, different for different 
processes, see Ref.~\cite{MV4}, in which the $N^{\,0}$ terms and their 
soft-gluon exponentiation can be used as a quantitative substitute for the full
coefficient functions.

Consequently several groups have addressed this issue with various approaches 
in the past few years, for work by others see 
Refs.~\cite{Laenen:2008ux,Laenen:2008gt,Gardi:2010rn,Laenen:2010uz} and 
\cite{Grunberg:2007nc,Grunberg:2009yi,Grunberg:2009vs}. 
However, to the best of our knowledge, this research has not yet led to 
explicit all-order predictions for complete and exact coefficients of the 
next-to-leading and next-to-next-to-leading logarithms, or even the leading 
logarithms to quantities such as the gluon contribution to most important 
structure function $\Ftwo$.

In Refs.~\cite{MV3,MV5} a resummation of the highest three $1/N$-suppressed 
logarithms (actually the $1/N$ behaviour is more formal than real for 
practically relevant values of $N$) for various quark coefficient functions has
been obtained by studying the large-$N$ behaviour of non-singlet physical 
evolution kernels. 
These quantities express the scaling violations of a physical quantity in terms
of the same physical quantity and hence do not depend on the scheme for the
factorization of the mass singularities. It turns out that the non-singlet
physical kernels show an only single-logarithmic large-$N$ enhancement also
beyond the dominant $N^{\,0}$ contributions (where this feature is a simple
consequence of the soft-gluon exponentiation, see Ref.~\cite{NV4}) up to at
least the next-to-next-to-leading or next-to-next-to-next-to-leading orders 
(NNLO or N$^3$LO). The results of Refs.~\cite{MV3,MV5} are based on the 
conjecture that this behaviour continues to all orders in the strong coupling 
$\as$.

Completely analogous observations were made \cite{SMVV1,LL2010} for the 
physical-kernel matrices of the systems ($\Ftwo,F_\phi$) and ($\Ftwo,\FL$) of 
flavour-singlet structure functions~\cite{FP82,Catani96,BRvN00}.
Unlike in the non-singlet case, recall Eq.~(\ref{Pii-xto1}), the singlet 
kernels receive double-logarithmic contributions from both the splitting
functions (\ref{Pij-xto1}) and the coefficient functions in Eqs.\
(\ref{Cak-xto1}) and (\ref{CLk-xto1}). Consequently the single-logarithmic
enhancement of the physical kernels can only provide one all-order constraint 
between two quantities. Definite fourth-order predictions of the highest 
three logarithms are possible though: first for the N$^3$LO splitting functions
\cite{SMVV1} using the diagram calculations of the corresponding three-loop
coefficient functions \cite{MVV5,MVV6,SMVV1}, and then, using these 
predictions, for the N$^3$LO (fourth-order) coefficient functions for $\FL$ 
\cite{LL2010}.

While we expect the results of Refs.~\cite{MV5,MV3,SMVV1,LL2010} to be the
final word on the predicted coefficients, clearly more work is required to put
them on a firmer theoretical footing and to extend them to all orders also for
the off-diagonal splitting functions and flavour-singlet coefficient functions.
At the leading-logarithmic level this was done in Ref.~\cite{AV2010} by 
finding the now expected all-order iterative structure of the unfactorized 
partonic structure functions (forward-Compton amplitudes). In the present paper
we have extended these results to the next-to-leading and next-to-next-to-%
leading (NLL and NNLL) logarithms. Note that this is the counting of a 
resummation, not that of a more powerful exponentiation 
\cite{SGlue1,SGlue2,SGlue3,SGlue4,SGlue5}: the present predictive power in 
terms of higher-order coefficients corresponds to that of a next-to-leading 
logarithmic exponentiation, see, e.g., Ref.~\cite{MVV7}.

In order to achieve this, we have employed two independent methods. The first
is a direct generalization of the amplitude iteration of Ref.~\cite{AV2010} to
higher logarithmic accuracy, presented at NLL accuracy in 
Eqs.~(\ref{eq:NLLit}) -- (\ref{eq:gphiq2g}) above. The second, worked out to
NNLL accuracy in the rest of Section 3 and Section 4, is conceptually extremely 
simple and revealing but not iterative, since all coefficients are determined 
order-by-order in $\as$ without any explicit reference to complete lower-order
amplitudes. This method is based on only the form of the $D$-dimensional phase 
space for inclusive DIS and the all-order mass-factorization formula 
(guaranteed by the operator-product expansion) for the unfactorized structure 
functions. In this way we have been able to verify all DIS predictions in 
Refs.~\cite{MV5,MV3,SMVV1,LL2010}, extend them to the fourth logarithms for
$F_{\,2,\rm ns}$ and to all orders in $\as$ for the highest three logarithms 
of the off-diagonal splitting functions and the coefficient functions 
$C_{2,\rm g}$, $C_{L,\rm g}$ and $C_{\phi,\rm q}$. Our results prove the 
non-singlet conjecture of Refs.~\cite{MV5,MV3} and show that also the 
above singlet physical evolution kernels are single-log enhanced to all orders. 

The resulting all-order off-diagonal splitting functions and the above 
coefficient functions can be written down in a closed all-order form to NLL 
(NNLL for $C_{L,\rm g}$) accuracy in terms of the apparently new special 
functions introduced in Ref.~\cite{AV2010}.
We did not find (so far) similar relations for all colour factors of 
$P_{\rm qg}$ and $P_{\rm gq}$ and for $C_{2,\rm g}$ and $C_{\phi,\rm q}$ at the
NNLL level, but can analytically determine their coefficients to `any' order 
in $\as$. The splitting-function results show a particular colour structure 
with all double logarithmic contributions, $\as^{\,n}\ln^{\,\ell} \x1$ with 
$\ell \geq n$, vanishing in the supersymmetric limit $\cf=\ca$. This pattern
does not hold for the more complicated coefficient functions which however can 
be expressed in terms of exponentials in the limit.

Numerically the resummation corrections to the $\as^{\,3}$ splitting functions
appear to be rather small and quickly converging: at the present level of
accuracy only the fourth-order terms have an  impact. The corresponding 
corrections for $C_{L,\rm q}$ are larger and receive
relevant fourth- and fifth-order contributions, but they are still small
compared to those up to order $\as^{\,6}$ for $C_{2,\rm g}$ and especially 
$C_{L,\rm g}$. Taking into account that the highest three logarithms most
likely underestimate the corrections at high orders of $\as$, these results
reinforce the third-order findings of Ref.~\cite{MVV6} which indicated that the
perturbative expansion of $C_{L,\rm g}$ is not well-behaved even at moderately
large $N$ in the important region of scales $\Qs \approx 25 \ldots 50 
\mbox{ GeV}^2$ corresponding to $\as(\Qs) \approx\, 0.2$.

Beyond first estimates of the numerical impact of higher orders at large $N$,
our results will prove very useful once the fixed-moment calculations of
Refs.~\cite{Larin:1994vu,Larin:1997wd,Retey:2000nq} have been extended to the
fourth order in $\as$. First results have already been presented on four-loop 
second moments and sum rules \cite{Baikov:2006ai,Baikov:2010je}, hence this 
extension can be expected in the foreseeable future. The main way to use a 
couple of moments of, e.g., the N$^3$LO splitting functions will be by 
effective $x$-space parametrizations analogous to those of Refs.~\cite{NV2,NV3}
at NNLO. Already knowing the coefficient of $\ln^{\,\ell} \x1$ for $\ell = 
4,\,5,\,6$ will then considerably assist achieving a decent accuracy from a
limited number of moments. The situation is analogous for the coefficient
functions.

There appears to be no reason why our determination of the highest double 
logarithms cannot be extended beyond the $N^{\,-1}$ ($N^{\,-2}$ for 
$C_{L,\rm g}$) terms. An all-order calculation would obviously be very 
cumbersome given the form of some of the intermediate expression in the present
limit, but a fourth-order calculation would definitely be feasible. We would
expect to recover the results of Refs.~\cite{MV5,SMVV1} for the coefficient
of the highest three logarithms at all-powers of $\x1$ of $c_{a,\rm q}^{\,(4)}$
with $a = 2,\,3,\,L$, $P_{\rm qg}^{\,(3)}$ and $P_{\rm gq}^{\,(3)} $ in this 
manner, and to derive corresponding new results for $c_{a,\rm g}^{\,(4)}$ 
($a = 2,\,\,L$) and $c_{\phi,\rm q}^{\,(4)}$. The main application of such 
results would be as a check of an all-$x$ fourth-order calculation of the
splitting functions and DIS coefficient functions which, however, we do not
expect for the near future.

Due to the similar phase-space integrations \cite{RvN1,RvN2}, the present 
approach can be carried over directly to the case of semi-inclusive $e^+e^-$
annihilation \cite{PDG2010} (but not, unfortunately, to the Drell-Yan process 
and inclusive Higgs production in $pp$ collisions). Only the diagonal NNLO 
`timelike' splitting functions for parton fragmentation have been determined up
to now \cite{PTqq2,PTgg2}, hence a NNLL resummation is not yet possible in this
case as the third logarithms of the NNLO splitting functions are a necessary
input. 
Therefore this case will be addressed in a future publication after the 
determination of the off-diagonal timelike splitting functions \cite{AMVprp}.

Let us finally stress again that the present double-logarithmic resummation is
not relying on any specific large-$N$ structure beyond the general form of the
phase space integrations. One may hope that, as in the case of the soft-gluon 
exponentiation of the $N^{\,0}$ coefficient functions, such additional 
structures can be found, e.g., using the new approach of 
Refs.~\cite{Laenen:2008gt,Gardi:2010rn} or by improved applications of the
soft-collinear effective theory to DIS, see, e.g., 
Refs.~\cite{SCET1,SCET2,SCET3,SCET4}. It should then become possible to resum 
also lower logarithms analogous to the standard threshold resummation.
%
\subsection*{Acknowledgements}

The work of G.S. and A.A. has been funded by the UK Science \& Technology 
Facilities Council (STFC), in the latter case under grant number ST/G00062X/1.
Also the research of A.V., who is grateful to D. Broadhurst and S. Moch 
for useful discussions, has been supported by this grant.

\newpage

%
\renewcommand{\theequation}{A.\arabic{equation}}
\setcounter{equation}{0}
\section*{Appendix A: Large-$N$ splitting functions at NNLL accuracy}
\label{sec:AppA}
Here we provide the expansion coefficients of the highest three logarithms
of the off-diagonal splitting functions, defined via
\bea
 N{\widetilde{P}}^{\,(n)}_{ij}(N) &\!=\!& 
 \;\cafn \ln^{\,2n\,} \Ntil\; \frac{4^n}{(n!)^2}\: D^{\,(n)}_{\rm LL} 
 \; + \; 
 \cafn \ln^{\,2n-1} \Ntil\; \frac{4^n}{(n!)^2}\: D^{\,(n)}_{\rm NLL,1} 
\nn \\ & & \mbox{\hspn} + \;
 \cafnm \ln^{2n-1}N \:\frac{4^{n-1}}{n!(n-1)!}\, 
 \left[ \cf D^{\,(n)}_{\rm NLL,2} + \b0 D^{\,(n)}_{\rm NLL,3} \right] 
\nn \\ & & \mbox{\hspn} + \;
 \cafn \ln^{\,2n-2} \Ntil\; \frac{4^n}{(n!)^2}\: D^{(n)}_{\,\rm NNL,1} 
\nn \\ & & \mbox{\hspn} + \;
 \cafnm \ln^{\,2n-2} \Ntil\; \frac{4^{n-1}}{n!(n-1)!}\, 
 \left[ \cf D^{\,(n)}_{\rm NNL,2} + \b0 D^{\,(n)}_{\rm NNL,3} \right] 
\nn \\ & & \mbox{\hspn} + \;
 \cafnmm \ln^{\,2n-2} \Ntil\; \frac{4^{n-2}}{n!(n-2)!}\,
 \left[ \cfs D^{\,(n)}_{\rm NNL,4} + \cf\b0 D^{\,(n)}_{\rm NNL,5} 
 + \bb02 D^{\,(n)}_{\rm NNL,6} \right] 
\nn \\ & & \mbox{\hspn} + \;
 {\cal O}(\ln^{\,2n-3} \Ntil)
\label{eq:pNNL} 
\eea
with $\widetilde{P}_{\rm qg}(N) = P_{\rm qg}/\nf$, 
$\widetilde{P}_{\rm gq}(N) = P_{\rm gq}/\cf$ and $\caf = \ca - \cf$. It is
understood that the coefficients on the r.h.s.~depend on the splitting
function under consideration. $D_{\rm NLL,1}$ vanishes for $P_{\rm qg}$, but 
not for $P_{\rm gq}$. 
The coefficients of the leading and next-to-leading logarithms, for which 
closed expressions have been written down in Eqs.~(\ref{eq:pqgNL}) and 
(\ref{eq:pgqNL}), are given up to order $\as^{\,18}$ in Table 1. 
Their counterparts for the third logarithms are provided in Table 2 for 
$P_{\rm qg}$ and Table 3 for $P_{\rm gq}$. The closed expression leading to 
most of these coefficients is not known yet. 

\begin{table}[p]
\vspace*{1mm}
\begin{center}
\small
\begin{tabular}{|c|cccc|ccc|}\hline
 & & & & & & & \\[-3mm]
 &\multicolumn{4}{c|}{$P_{\rm gq}^{\,(n)}$} 
 &\multicolumn{3}{c|}{$P_{\rm qg}^{\,(n)}$}\\[2mm] 
  \hline
 & & & & & & & \\[-3mm]
 $n$ & $D^{\,(n)}_{\rm LL}$    & $D^{\,(n)}_{\rm NLL,1}$ 
     & $D^{\,(n)}_{\rm NLL,2}$ & $D^{\,(n)}_{\rm NLL,3}$ 
     & $D^{\,(n)}_{\rm LL}$    & $D^{\,(n)}_{\rm NLL,2}$ 
     & $D^{\,(n)}_{\rm NLL,3}$\\[2mm] \hline
 & & & & & & & \\[-2mm]
$0$ & $2$ & $0$ & $0$ & $0$ & $2$  & $0$ & $0$\\[2mm]
$1$ & $1$ & $-2\phm$ & $12$ & $-4\phm$ & $-1\phm$  & $0$ & $0$ \\[2.5mm]
$2$ & $\ds\frac{1}{3}$ & $-2\phm$ & $5$ & $-\ds\frac{5}{3}\phm$ & 
$\ds\frac{1}{3}$ & $-1\phm$ & $\ds\frac{1}{3}$\\[3.5mm]
$3$ & $0$ & $-1\phm$ & $1$ & $-\ds\frac{1}{6}\phm$ & $0$ & $1$ & 
$-\ds\frac{1}{6}\phm$\\[3.5mm]
$4$ & $-\ds\frac{1}{15}\phm$ & $0$ & $-\ds\frac{2}{5}\phm$ & 
$\ds\frac{3}{10}$ & $-\ds\frac{1}{15}\phm$ & $-\ds\frac{2}{5}\phm$ & 
$-\ds\frac{1}{30}\phm$\\[3.5mm]
$5$ & $0$ & $\ds\frac{1}{3}$ & $-\ds\frac{1}{5}\phm$ & $\ds\frac{1}{30}$ & 
$0$ & $-\ds\frac{1}{5}\phm$ & $\ds\frac{1}{30}$\\[3.5mm]
$6$ & $\ds\frac{1}{21}$ & $0$  & $\ds\frac{2}{7}$ & $-\ds\frac{11}{42}\phm$ & 
$\ds\frac{1}{21}$ & $\ds\frac{2}{7}$ & $\ds\frac{1}{14}$\\[3.5mm]
$7$ & $0$  & $-\ds\frac{1}{3}\phm$ & $\ds\frac{1}{7}$ & 
$-\ds\frac{1}{42}\phm$ & $0$ & $\ds\frac{1}{7}$ & $-\ds\frac{1}{42}\phm$
\\[3.5mm]
$8$ & $-\ds\frac{1}{15}\phm$ & $0$  & $-\ds\frac{2}{5}\phm$ & 
$\ds\frac{13}{30}$ & $-\ds\frac{1}{15}\phm$ & $-\ds\frac{2}{5}\phm$ & 
$-\ds\frac{1}{6}\phm$\\[3.5mm]
$9$ & $0$ & $\ds\frac{3}{5}$ & $-\ds\frac{1}{5}\phm$ & $\ds\frac{1}{30}$ & 
$0$ & $-\ds\frac{1}{5}\phm$ & $\ds\frac{1}{30}$\\[3.5mm]
$10$ & $\ds\frac{5}{33}$ & $0$ & $\ds\frac{10}{11}$ & 
$-\ds\frac{25}{22}\phm$ & $\ds\frac{5}{33}$ & $\ds\frac{10}{11}$ & 
$\ds\frac{35}{66}$ \\[3.5mm]
$11$ & $0$ & $-\ds\frac{5}{3}\phm$ & $\ds\frac{5}{11}$ & 
$-\ds\frac{5}{66}\phm$ & $0$ & $\ds\frac{5}{11}$ & $-\ds\frac{5}{66}\phm$ 
\\[3.5mm]
$12$ & $-\ds\frac{691}{1365}\phm$ & $0$ & $-\ds\frac{1382}{455}\phm$ & 
$\ds\frac{11747}{2730}$ & $-\ds\frac{691}{1365}\phm$ & 
$-\ds\frac{1382}{455}\phm$ & $-\ds\frac{2073}{910}\phm$\\[3.5mm]
$13$ & $0$ & $\ds\frac{691}{105}$ & $-\ds\frac{691}{455}\phm$  & 
$\ds\frac{691}{2730}$ & $0$ & $-\ds\frac{691}{455}\phm$  & 
$\ds\frac{691}{2730}$\\[3.5mm]
$14$ & $\ds\frac{7}{3}$ & $0$ & $14$ & $-\ds\frac{133}{6}\phm$ & 
$\ds\frac{7}{3}$ & $14$ & $\ds\frac{77}{6}$ \\[3.5mm]
$15$ & $0$ & $-35\phm$ & $7$ & $-\ds\frac{7}{6}\phm$ & $0$ & $7$ & 
$-\ds\frac{7}{6}\phm$\\[3.5mm]
$16$ & $\;-\ds\frac{3617}{255}\phm$ & $0$ & $-\ds\frac{7234}{85}\phm$ & 
$\ds\frac{25319}{170}\;$ & $\;-\ds\frac{3617}{255}\phm$ & 
$-\ds\frac{7234}{85}\phm$ & $-\ds\frac{47021}{510}\phm\;$\\[3.5mm]
$17$ & $0$ & $\ds\frac{3617}{15}$ & $-\ds\frac{3617}{85}\phm$ & 
$\ds\frac{3617}{510}$ & $0$ & $-\ds\frac{3617}{85}\phm$ & 
$\ds\frac{3617}{510}$ \\[3.5mm]
\hline
\end{tabular} 
\vspace{2mm} 
\caption{The coefficients of the leading and next-to-leading large-$N$ 
logarithms of the off-diagonal splitting functions $P_{\rm gq}$ and 
$P_{\rm qg}$ as defined in Eq.~(\ref{eq:pNNL}) up to the 18-th order 
$\ar=\as/(4\pi)$. Note the respective appearance of the numerators 691 and 
3617 from $n=12$ and $n=16$ which clearly signals the presence of the
Bernoulli numbers (\ref{Bnumbers}) also in the NLL coefficients.
\label{tab:pNL}}   
\end{center}
\vspace*{-1mm}
\end{table} 

\begin{table}[p]
\begin{center}
\small
\begin{tabular}{|c|cccccc|}\hline
 & & & & & & \\[-3mm]
 $n$ &
 $D^{\,(n)}_{\rm NNL,1}$ & $D^{\,(n)}_{\rm NNL,2}$ &
 $D^{\,(n)}_{\rm NNL,3}$ & $D^{\,(n)}_{\rm NNL,4}$ &
 $D^{\,(n)}_{\rm NNL,5}$ & $D^{\,(n)}_{\rm NNL,6}$ \\[2mm] \hline
 & & & & & & \\[-2mm]
$0$ & $0$ & $0$ & $0$ & $0$ & $0$ & $0$\\[2.5mm]
$1$ & $1$ & $14-4\z2$ & $0$ & $0$ & $0$ & $0$\\[4mm]
$2$ & $-\ds\frac{4}{3}+4\z2\phm$ & $-\ds\frac{50}{3}+8\z2\phm$ & 
$-\ds\frac{10}{3}\phm$ & $0$ & $0$ & $0$\\[4mm]
$3$ & $\ds\frac{5}{6}-5\z2$ & $\ds\frac{121}{9}-6\z2$ & $\ds\frac{20}{9}$ & 
$-\ds\frac{3}{4}\phm$ & $\ds\frac{3}{4}$ & $-\ds\frac{1}{6}\phm$\\[4mm]
$4$ & $\ds\frac{2}{3}+3\z2$ & $-\ds\frac{19}{3}+\z2\phm$ & $\ds\frac{5}{6}$ & 
$\ds\frac{8}{5}$ & $-\ds\frac{11}{5}\phm$ & $\ds\frac{1}{15}$\\[4mm]
$5$ & $-\ds\frac{3}{2}+3\z2\phm$ & $-\ds\frac{1}{5}+2\z2\phm$ & $-2\phm$ & 
$-\ds\frac{63}{40}\phm$ & $-\ds\frac{3}{40}\phm$ & $\ds\frac{1}{20}$\\[4mm]
$6$ & $-\ds\frac{1}{6}(2+3\z2)\phm$ & $\ds\frac{1}{9}(19-3\z2)$ & 
$-\ds\frac{5}{18}\phm$ & $\ds\frac{12}{35}$ & $\ds\frac{4}{7}$ & 
$\ds\frac{13}{1260}$\\[4mm]
$7$ & $\ds\frac{1}{6}(17-33\z2)$ & $\ds\frac{3}{7}-\ds\frac{8}{3}\z2$ & 
$\ds\frac{55}{21}$ & $\ds\frac{31}{28}$ & $\ds\frac{17}{84}$ & 
$-\ds\frac{5}{84}\phm$\\[4mm]
$8$ & $\ds\frac{2}{9}(2+3\z2)$ & $-\ds\frac{1}{9}(19-3\z2)\phm$ & 
$\ds\frac{5}{18}$ & $-\ds\frac{6}{7}\phm$ & $-\ds\frac{8}{7}\phm$ & 
$-\ds\frac{46}{315}\phm$\\[4mm]
$9$ & $-\ds\frac{83}{10}+\ds\frac{78}{5}\z2\phm$ & 
$-\ds\frac{61}{45}+6\z2\phm$ & $-\ds\frac{52}{9}\phm$ &
 $-\ds\frac{123}{80}\phm$ & $-\ds\frac{39}{80}\phm$ & $\ds\frac{7}{60}$\\[4mm]
$10$ & $-\ds\frac{1}{2}(2+3\z2)\phm$ & $\ds\frac{1}{5}(19-3\z2)$ &
$-\ds\frac{1}{2}\phm$ & $\ds\frac{376}{165}$ & $\ds\frac{1712}{495}$ & 
$\ds\frac{171}{220}$\\[4mm]
$11$ & $\ds\frac{5}{6}(41-75\z2)$ & $\ds\frac{185}{33}-20\z2$ &
$\ds\frac{625}{33}$ & $\ds\frac{153}{44}$ & $\ds\frac{69}{44}$ & 
$-\ds\frac{15}{44}\phm$\\[4mm]
$12$ & $\ds\frac{5}{3}(2+3\z2)$ & $-\ds\frac{5}{9}(19-3\z2)\phm$ & 
$\ds\frac{25}{18}$ & $-\ds\frac{3714}{455}\phm$ & 
$-\ds\frac{72656}{5005}\phm$ & $\hsps-\ds\frac{413117}{90090}\phm\hsps$\\[4mm]
$13$ & $\hsps-\ds\frac{691}{70}(19-34\z2)\phm\hsps$ & 
$\hsps-\ds\frac{691}{1365}(59-182\z2)\phm\hsps$ & 
$\hsps-\ds\frac{23494}{273}$ & $\hsps-\ds\frac{42151}{3640}\phm\hsps$ & 
$\hsps-\ds\frac{73937}{10920}\phm\hsps$ & $\hsps\ds\frac{7601}{5460}\hsps$
\\[4mm]
$14$ & $-\ds\frac{691}{90}(2+3\z2)\phm$ & $\ds\frac{691}{315}(19-3\z2)$ & 
$-\ds\frac{691}{126}\phm$ & $\ds\frac{17772}{455}$ & 
$\ds\frac{36728}{455}$ & $\ds\frac{534389}{16380}$\\[4mm]
$15$ & $\ds\frac{35}{6}(227-399\z2)$ & $\ds\frac{1813}{9}-560\z2$ & 
$\ds\frac{4655}{9}$ & $\ds\frac{213}{4}$ & $\ds\frac{153}{4}$ & 
$-\ds\frac{91}{12}\phm$\\[4mm]
$16$ & $\ds\frac{140}{3}(2+3\z2)$ & $-\ds\frac{35}{3}(19-3\z2)\phm$ & 
$\ds\frac{175}{6}$ & $-\ds\frac{103202}{425}\phm$ & 
$-\ds\frac{733288}{1275}\phm$ & $-\ds\frac{71989}{255}\phm$\\[4mm]

$17$ & $\;\hsps-\ds\frac{3617}{30}(97-168\z2)\phm\hsps\;$ & 
$\hsps-\ds\frac{3617}{15}(7-18\z2)\phm$ & 
$\hsps-\ds\frac{202552}{51}\hsps\phm$ & 
$\hsps-\ds\frac{878931}{2720}\phm$ & 
$\hsps-\ds\frac{748719}{2720}\hsps\phm$ & 
$\hsps\ds\frac{3617}{68}\hsps$\\[4mm]
\hline
\end{tabular}
\vspace{2mm}
\caption{The coefficients of the next-to-next-to-leading large-$N$ logarithms
of the off-diagonal splitting function $P_{\rm qg}$, as defined in
Eq.~(\ref{eq:pNNL}), up to the 18-th order $\ar=\as/(4\pi)$.
\label{tab:pqgNNL}}
\end{center}
\end{table}

\begin{table}[p]
\begin{center}
\small
\begin{tabular}{|c|cccccc|}\hline
 & & & & & & \\[-3mm]
 $n$ & 
 $D^{\,(n)}_{\rm NNL,1}$ & $D^{\,(n)}_{\rm NNL,2}$ & 
 $D^{\,(n)}_{\rm NNL,3}$ & $D^{\,(n)}_{\rm NNL,4}$ & 
 $D^{\,(n)}_{\rm NNL,5}$ & $D^{\,(n)}_{\rm NNL,6}$ \\[2mm] \hline
 & & & & & & \\[-2mm]
$0$ & $0$ & $0$ & $0$ & $0$ & $0$ & $0$\\[2.5mm]
$1$ & $\ds\frac{7}{3}$ & $-\ds\frac{44}{3}-4\z2\phm$ & $\ds\frac{32}{3}$ & 
 $0$ & $0$ & $0$\\[4mm]
$2$ & $\ds\frac{4}{3}$ & $\ds\frac{5}{3}-8\z2$ & $\ds\frac{31}{3}$ & $54$ & 
$-36\phm$ & $6$\\[4mm]
$3$ & $\ds\frac{1}{6}+\z2$ & $\ds\frac{56}{9}-6\z2$ & $\ds\frac{52}{9}$ & 
$\ds\frac{75}{4}$ & $-\ds\frac{41}{4}$\phm & $\ds\frac{4}{3}$\\[4mm]
$4$ & $-\ds\frac{1}{3}(2-9\z2)\phm$ & $\ds\frac{1}{3}(10-3\z2)$ & 
$-\ds\frac{1}{3}\phm$ & $\ds\frac{8}{5}$ & $\ds\frac{8}{5}$ & 
$-\ds\frac{32}{45}\phm$\\[4mm]
$5$ & $-\ds\frac{5}{6}+3\z2\phm$ & $-\ds\frac{2}{3}+2\z2\phm$ & 
$-\ds\frac{44}{15}\phm$ & $-\ds\frac{81}{40}\phm$ & $\ds\frac{81}{40}$ & 
$-\ds\frac{3}{10}\phm$\\[4mm]
$6$ & $\ds\frac{1}{6}(2-9\z2)$ & $-\ds\frac{1}{9}(10-3\z2)\phm$ & 
$\ds\frac{1}{9}$ & $\ds\frac{12}{35}$ & $-\ds\frac{57}{35}\phm$ & 
$\ds\frac{937}{1260}$\\[4mm]
$7$ & $\ds\frac{1}{18}(31-99\z2)$ & $\ds\frac{80}{63}-\ds\frac{8}{3}\z2$ & 
$\ds\frac{34}{9}$ & $\ds\frac{41}{28}$ & $-\ds\frac{45}{28}\phm$ &
 $\ds\frac{61}{252}$\\[4mm]
$8$ & $-\ds\frac{2}{9}(2-9\z2)\phm$ & $\ds\frac{1}{9}(10-3\z2)$ &
$-\ds\frac{1}{9}\phm$ & $-\ds\frac{6}{7}\phm$ & $\ds\frac{19}{7}$ & 
$-\ds\frac{451}{315}\phm$\\[4mm]
$9$ & $-\ds\frac{11}{2}+\ds\frac{78}{5}\z2\phm$ & 
$-\ds\frac{146}{45}+6\z2\phm$ & $-\ds\frac{376}{45}\phm$ & 
$-\ds\frac{33}{16}\phm$ & $\ds\frac{197}{80}$ & $-\ds\frac{3}{8}\phm$\\[4mm]
$10$ & $\ds\frac{1}{2}(2-9\z2)$ & $-\ds\frac{1}{5}(10-3\z2)\phm$ &
$\ds\frac{1}{5}$ & $\ds\frac{376}{165}$ & $-\ds\frac{1171}{165}\phm$ & 
$\ds\frac{25517}{5940}$\\[4mm]
$11$ & $\ds\frac{5}{6}(29-75\z2)$ & $\ds\frac{380}{33}-20\z2$ & 
$\ds\frac{910}{33}$ & $\ds\frac{207}{44}$ & $-\ds\frac{267}{44}\phm$ & 
$\ds\frac{41}{44}$\\[4mm]
$12$ & $-\ds\frac{5}{3}(2-9\z2)\phm$ & $\ds\frac{5}{9}(10-3\z2)$ & 
$-\ds\frac{5}{9}\phm$ & $-\ds\frac{3714}{455}\phm$ & 
$\ds\frac{133953}{5005}$ & 
$\hsps-\ds\frac{1652771}{90090}\phm\hsps$\\[4mm]
$13$ & $\;\hsps-\ds\frac{691}{630}(127-306\z2)\phm\hsps\;$ & 
$\hsps-\ds\frac{1382}{4095}(163-273\z2)\phm\hsps$ & 
$\hsps-\ds\frac{516868}{4095}\phm\hsps$ & 
$\hsps-\ds\frac{57353}{3640}\phm\hsps$ & 
$\hsps\ds\frac{15893}{728}\hsps$ & 
$\hsps-\ds\frac{2764}{819}\phm\hsps$\\[4mm]
$14$ & $\ds\frac{691}{90}(2-9\z2)$ & $-\ds\frac{691}{315}(10-3\z2)\phm$ 
& $\ds\frac{691}{315}$ & $\ds\frac{17772}{455}$ & 
$-\ds\frac{62497}{455}\phm$ & $\ds\frac{1725089}{16380}$\\[4mm]
$15$ & $\ds\frac{245}{6}(25-57\z2)$ & $\ds\frac{3080}{9}-560\z2$ & 
$\ds\frac{6874}{9}$ & $\ds\frac{291}{4}$ & $-\ds\frac{431}{4}\phm$ & 
$\ds\frac{67}{4}$\\[4mm]
$16$ & $-\ds\frac{140}{3}(2-9\z2)\phm$ & $\ds\frac{35}{3}(10-3\z2)$ & 
$-\ds\frac{35}{3}\phm$ & $-\ds\frac{103202}{425}\phm$ & 
$\ds\frac{390957}{425}$ & 
$\hsps-\ds\frac{2985994}{3825}\phm\hsps$\\[4mm]
$17$ & $\;\hsps-\ds\frac{25319}{30}(11-24\z2)\phm\hsps\;$ & 
$\hsps-\ds\frac{7234}{255}(95-153\z2)\phm\hsps$ &
$\hsps-\ds\frac{1504672}{255}\phm\hsps$ & 
$\hsps-\ds\frac{1204461}{2720}\phm\hsps$ & 
$\hsps\ds\frac{379785}{544}\hsps$ & 
$\hsps-\ds\frac{148297}{1360}\phm\hsps$\\[4mm]
\hline
\end{tabular}
\vspace{2mm}  
\caption{As Table 2, but for the splitting function $P_{\rm gq}$. 
\label{tab:pgqNNL}}   
\end{center}
\end{table} 

%
\renewcommand{\theequation}{B.\arabic{equation}}
\setcounter{equation}{0}
\section*{Appendix B: Large-$N$ coefficient functions at NNLL accuracy}
\label{sec:AppB}
Finally we turn to the NNLL contributions to the off-diagonal coefficient 
functions $C_{2,\rm g}$ and $C_{\phi,\rm q}$. The more complicated colour
structure, with $3n-3$ contributions at order $n$, precludes a representation 
in the form of Tables 2 and 3. 
Instead we provide the resulting expression at order $\as^{\,4}$ and 
$\as^{\,5}$ for general SU(N) colour factors, and then the six-figure values 
for QCD to order $\as^{\,12}$ which are more than sufficient for numerical
applications. In all cases we include also the LL and NLL terms, for which 
closed expression have been given in Eqs.~(\ref{eq:c2gNL}) and 
Eqs.~(\ref{eq:cphiqNL}) above.

The fourth- and fifth-order contributions to the gluon coefficient function
for $\Ftwo$ are given by
\bea
 N\:\!c_{2,\rm g}^{\,(4)}(N) &\!=\!\!& \mbox{}
 -\,\nf\ln^7 \Ntil\:
 \bigg[\, {8 \over 3}\,\cft+{4 \over 3}\,\caf\cfs+{4 \over 3}\,\cafs\cf
    +{46 \over 135}\, \caft\bigg] 
\nn \\[1.5mm] & & \mbox{}
 -\,\nf\ln^6 \Ntil\:
 \bigg[\,{8 \over 3}\,\cfs\,\b0+{62 \over 3}\,\cft
    +\caf \bigg(\, {11 \over 9}\,\cf\b0 +{49 \over 3}\,\cfs \bigg) 
\nn \\[1mm] & & \mbox{}
 \quad+\,\cafs \bigg(\, {56 \over 135}\,\b0+{1061 \over 90}\,\cf \bigg)
    +{8 \over 3}\,\caft \bigg] 
\nn \\[1.5mm] & & \mbox{}
 -\,\nf\ln^5 \Ntil\:
 \bigg[\, {2 \over  3}\,\cf\bb02+30\,\cfs\,\b0+{230 \over 3}\,\cft
    -40\,\z2\,\cft 
\nn \\[1mm] & & \mbox{}
 \quad+\,\caf \bigg(\, {23 \over 180}\,\bb02+{15761 \over 1080}\,\cf\b0
    +{122647 \over 1080}\,\cfs-{164 \over 3}\:\z2\,\cfs \bigg) 
\nn \\[1mm] & & \mbox{} 
 \quad+\,\cafs \bigg(\, {259 \over 54}\,\b0+{2423 \over 27}\,\cf
    -33\,\z2\,\cf \bigg) +\caft \bigg(\, {730 \over 27}-{97 \over 9}\,\z2 
    \bigg) \bigg] \qquad
\nn \\[1.5mm] & & \mbox{}
 +\,{\cal O}(\ln^4 \Ntil\,)
\label{eq:C2g4}
\eea
and
\bea
 N\:\!c_{2,\rm g}^{\,(5)}(N) &\!=\!\! & \mbox{}
 -\,\nf\ln^9 \Ntil\:
 \bigg[\, {4 \over 3}\,\cff+{8 \over 9}\,\caf\cft
    +{4 \over 3}\,\cafs\cfs+{92 \over 135}\,\caft\cf+{2 \over 15}\,\caff\bigg] 
\nn \\[1.5mm] & & \mbox{}
 -\,\nf\ln^8 \Ntil\:
 \bigg[\, {8 \over 3}\cft\,\b0+{40 \over 3}\,\cff
    +\caf \bigg(\, {5 \over 3}\,\cfs\,\b0+{41 \over 3}\,\cft \bigg) 
\nn \\[1mm] & & \mbox{}
 \quad+\,\cafs \bigg(\, {172 \over 135}\,\cf\b0+{701 \over 45}\,\cfs \bigg) 
 \,+\,\caft({419 \over 1350}\,\b0+{363 \over 50}\,\cf)
    +{4 \over 3}\,\caff \bigg] 
 \nn \\[1.5mm] & & \mbox{}
 -\,\nf\ln^7 \Ntil\:
 \bigg[{16 \over 9}\,\cfs\,\bb02+36\,\cft\,\b0+{206 \over 3}\,\cff
    -32\,\z2\,\cff 
 \nn \\[1mm] & & \mbox{}
 \quad+\,\caf \bigg(\, {199 \over 270}\,\cf\bb02+{14221 \over 540}\,\cfs\,\b0
    +{64007 \over 540}\,\cft-{160 \over 3}\,\z2\,\cft \bigg) 
 \nn \\[1mm] & & \mbox{}
 \quad+\,\cafs \bigg(\, {659 \over 2700}\,\bb02+{25097 \over 1350}\,\cf\b0
    +{123451 \over 900}\,\cfs-{466 \over 9}\,\z2\,\cfs \bigg) 
\nn \\[1mm] & & \mbox{}
 \quad+\,\caft \bigg(\, {1768 \over 405}\,\b0+{157504 \over 2025}\,\cf
    -{494 \over 15}\,\z2\,\cf \bigg) 
 \,+\,\caff \bigg(\, {36484 \over 2025}-{4976 \over 675}\,\z2 \bigg)
\bigg]
\nn \\[1mm] & & \mbox{}
 +\,{\cal O}(\ln^6 \Ntil\,) \; .
\label{eq:C2g5}
\eea
The corresponding results for the quark coefficient function for $F_\phi$ read
\bea
 N\:\!c_{\phi,\rm q}^{\,(4)}(N) &\!=\!\!& \mbox{}
 -\,\cf\ln^7 \Ntil\:
 \bigg[\, {8 \over 3}\,\cft+{20 \over 3}\,\caf\cfs+{20 \over 3}\,\cafs\cf
    +{314 \over 135}\,\caft\bigg] 
\nn \\[1.5mm] & & \mbox{}
-\,\cf\ln^6 \Ntil\:
 \bigg[\, {14 \over 3}\,\cfs\,\b0+{40 \over 3}\,\cft+\caf({67 \over 9}\,\cf\b0
    +{17 \over 3}\,\cfs) 
\nn \\[1mm] & & \mbox{}
 \quad+\,\cafs \bigg(\, {502 \over 135}\,\b0-{811 \over 90}\,\cf \bigg)
    -{68 \over 9}\,\caft\,\bigg] 
\nn \\[1.5mm] & & \mbox{}
 -\,\cf\ln^5 \Ntil\:
 \bigg[\,4\cf\bb02+{136 \over 3}\,\cfs\b0+{142 \over 3}\,\cft-32\,\z2\,\cft 
\nn \\[1mm] & & \mbox{}
 \quad+\,\caf \bigg(\, {187 \over 60}\,\bb02+{40427 \over 1080}\,\cf\b0
    +{60853 \over 1080}\,\cfs-{268 \over 3}\,\z2\,\cfs \bigg) 
\nn \\[1mm] & & \mbox{}
 \quad+\,\cafs \bigg(\, {301 \over 27}\,\b0+{452 \over 9}\,\cf
    -{497 \over 9}\,\z2\,\cf \bigg) 
 \,+\,\caft({398 \over 27}\,-{109 \over 9}\,\z2 )\bigg] \qquad
\nn \\[1mm] & & \mbox{}
 +\,{\cal O}(\ln^4 \Ntil\,)
\label{eq:Cphiq4}
\eea
and
\pagebreak
\bea
 N\:\!c_{\phi,\rm q}^{\,(5)}(N) &\!=\!\!& \mbox{}
 -\,\cf\ln^9 \Ntil\:
 \bigg[\, {4 \over 3}\,\cff+{40 \over 9}\,\caf\cft+{20 \over 3}\,\cafs\cfs
    +{628 \over 135}\,\caft\cf+{166 \over 135}\,\caff\bigg] 
\nn \\[1.5mm] & & \mbox{}
 -\,\cf\ln^8 \Ntil\:
 \bigg[\, 4\,\cft\,\b0+{26 \over 3}\,\cff 
    +\caf \bigg(\, {29 \over 3}\,\cfs\,\b0 +{29 \over 3}\,\cft \bigg) 
\nn \\[1mm] & & \mbox{}
 \quad+\,\cafs \bigg(\, ({1304 \over 135}\,\cf\b0-{17 \over 15}\,\cfs \bigg) 
 \,+\,\caft \bigg(\, {4601 \over 1350}\,\b0-{1519 \over 150}\,\cf)
    -{3296 \over 675}\,\caff \bigg)\bigg] 
\nn \\[1.5mm] & & \mbox{}
 -\,\cf\ln^7 \Ntil\:
 \bigg[\, {52 \over 9}\,\cfs\,\bb02+{386 \over 9}\,\cft\,\b0
    +{376 \over 9}\,\cff-{80 \over 3}\,\z2\,\cff 
\nn \\[1mm] & & \mbox{}
 \quad+\,\caf \bigg(\, {2453 \over 270}\,\cf\,\bb02
    +{31207 \over 540}\,\cfs\,\b0 +{41573 \over 540}\,\cft-96\,\z2\,\cft\bigg) 
\nn \\[1mm] & & \mbox{}
 \quad+\,\cafs \bigg(\, {2393 \over 540}\,\bb02+{33149 \over 900}\,\cf\b0
    +{12904 \over 135}\,\cfs-{326 \over 3}\,\z2\,\cfs \bigg) 
\nn \\[1mm] & & \mbox{}
 \quad+\,\caft \bigg(\, {4562 \over 675}\,\b0+{12254 \over 225}\,\cf
    -{916 \over 15}\,\z2\,\cf \bigg) 
 \,+\,\caff \bigg(\, {1648 \over 135}-{2638 \over 225}\,\z2 \bigg)\bigg] 
\nn \\[1mm] & & \mbox{}
 +\,{\cal O}(\ln^6 \Ntil\,) \; .
\label{eq:Cphiq5}
\eea
Our notation for the Tables 4-6, where we include $C_{L,\rm g}$ for the 
convenience of the reader, is
\bea
 N\,\widetilde{c}^{\,(n)}_{a, k}(N)\;\; &\!=\!\!& \mbox{}
 -\,\ln^{\,2n-1} \Ntil\: D^{\,(n)}_{a,\rm LL}
 \;-\; \ln^{\,2n-2} \Ntil\:
 \left[ D^{\,(n)1}_{a,\rm NLL} - D^{\,(n)2}_{a,\rm NLL}\,\nf \right] 
\nn \\[1mm] & & \mbox{}
 -\,\ln^{\,2n-3} \Ntil\:
 \left[ D^{\,(n)1}_{a,\rm NNL} - D^{\,(n)2}_{a,\rm NNL}\,\nf 
    + D^{\,(n)3}_{a,\rm NNL}\,\nfs \right] 
 \;+\; {\cal O}(\ln^{\,2n-4} \Ntil\,) \; , 
\label{eq:Cak-qcd}
\\[3mm]
 \nf^{\!\!-1}N^{\,2}\,c^{(n)}_{L,\rm g}(N)\!\! &\!=\!\!& \mbox{}
 +\,\ln^{\,2n-2} \Ntil\: D^{\,(n)}_{L,\rm LL}
 \;+\; \ln^{\,2n-3} \Ntil\:
 \left[ D^{\,(n)1}_{L,\rm NLL} - D^{\,(n)2}_{L,\rm NLL}\,\nf \right]
\nn \\[1mm] & & \mbox{}
 +\,\ln^{\,2n-4} \Ntil\:
 \left[ D^{\,(n)1}_{L,\rm NNL} - D^{\,(n)2}_{L,\rm NNL}\,\nf
    + D^{\,(n)3}_{L,\rm NNL}\,\flg11\nf
    + D^{\,(n)4}_{L,\rm NNL}\,\nfs \right]
\nn \\[1.5mm] & & \mbox{}
 +\; {\cal O}(\ln^{\,2n-5} \Ntil\,) 
\label{eq:CLg-qcd}
\eea 
with $\widetilde{C}_{2,\rm g} = C_{2,\rm g}/\nf$ and 
$\widetilde{C}_{\phi,\rm q} = C_{\phi,\rm q}/\cf$. $\flg11$ has been defined
in Eq.~(\ref{flg11}).

Finally it is worthwhile to note that a closed (if presumably rather lengthy) 
expression for the NNLL contributions to $C_{2,\rm g}$ and $C_{\phi,\rm q}$ 
can be derived once such an expression has been obtained for the corresponding
contributions to the splitting functions in Appendix A. At that point all 
quantities but $C_{2,\rm g}$ and $C_{\phi,\rm q}$ entering the vanishing 
off-diagonal NNLL elements of the physical evolution kernel
\beq
\label{F-evol}
  ~\frac{d F}{d \ln Q^2} \;=\; 
  \Big( \beta(\ar)\,\frac{d\, C}{d \ar}\:C^{\,-1} 
         + C\:\! P\:\! C^{\,-1} \Big) F \;\equiv\;  K F 
  \; = \; 0_{\,\rm NNLL}
\eeq
with the standard matrix $P$ of the singlet splitting functions and
\beq
  F \:=\:
  \Big( \begin{array}{c} \!\! F_2 \!\! \\ \! F_\phi \!\!
        \end{array} \Big)
\; , \;\;\;
  C \:=\:
  \Big( \! \begin{array}{cc} C_{2,\rm q}^{}\! & C_{2,\rm g}^{}\! \\
  C_{\phi,\rm q}\! & C_{\phi,\rm g}\! \end{array}\! \Big)
\; , \;\;\;
 K \:=\:
 \Big( \! \begin{array}{cc} K_{22}\! & K_{2\phi}\! \\
  K_{\phi 2}\! & K_{\phi\phi}\! \end{array}\! \Big)
\eeq
will be known, and Eqs.(\ref{F-evol}) can be solved for the NNLL parts of
$C_{2,\rm g}$ and $C_{\phi,\rm q}$. In fact, we have applied the analogous 
NLL procedure to find the closed expression given in Eqs.~(\ref{eq:c2gNL}) 
and (\ref{eq:cphiqNL}).

\begin{table}[p]
\vspace*{1mm}
\begin{center}
\small
\begin{tabular}{|c|cccccc|}\hline
 & & & & & & \\[-3mm]
 $n$ &
 $D^{\,(n)}_{2,\rm LL}$ & $D^{\,(n)1}_{2,\rm NL}$ &
 $D^{\,(n)2}_{2,\rm NL}$ & $D^{\,(n)1}_{2,\rm NNL}$ &
 $D^{\,(n)2}_{2,\rm NNL}$ & $D^{\,(n)3}_{2,\rm NNL}$ \\[2mm] \hline
 & & & & & & \\[-2mm]
$1$ & 2       & 2       & 0       & 0       & 0       & 0\\[2mm]
$2$ & 6.44444 & 24      & 0    & $-$3.43495 & 0       & 0\\[2mm]
$3$ & 11.9259 & 103.944 & 1.61728 & 285.481 & 13.7942 & 0\\[2mm]
$4$ & 16.7874 & 248.091 & 5.73937 & 1493.50 & 82.2181 & 0.489712\\[2mm]
$5$ & 19.5455 & 419.726 & 11.6100 & 4059.62 & 248.604 & 2.43393\\[2mm]
$6$ & 19.3641 & 561.667 & 17.4564 & 7676.98 & 508.882 & 6.25137\\[2mm]
$7$ & 16.5861 & 624.621 & 21.2105 & 11287.1 & 797.203 & 11.2984\\[2mm]
$8$ & 12.4589 & 592.811 & 21.5942 & 13619.9 & 1014.87 & 15.9871\\[2mm]
$9$ & 8.31829 & 489.124 & 18.8489 & 13915.5 & 1085.50 & 18.5769\\[2mm]
$10$& 4.99558 & 356.178 & 14.3660 & 12296.6 & 997.538 & 18.2401\\[2mm]
$11$& 2.72610 & 231.848 & 9.70788 & 9552.90 & 801.374 & 15.4587\\[2mm]
$12$& 1.36329 & 136.369 & 5.89075 & 6612.08 & 570.865 & 11.5051\\[2mm]
\hline
\end{tabular}
\vspace{2mm}
\caption{The LL, NLL and NNLL coefficients of $C_{2,\rm g}$ in QCD,
as defined in Eq.~(\ref{eq:Cak-qcd}), to the 12-th order in $\ar=\as/(4\pi)$.
All these coefficients further decrease at higher orders and tend to zero in
the infinite-order limit.
\label{tab:C2gNNLL}}
\end{center}
\end{table}
\begin{table}[p]
\vspace*{-1mm}
\begin{center}
\small
\begin{tabular}{|c|cccccc|}\hline
 & & & & & & \\[-3mm]
 $n$ &
 $D^{\,(n)}_{\phi,\rm LL}$ & $D^{\,(n)1}_{\phi,\rm NL}$ &
 $D^{\,(n)2}_{\phi,\rm NL}$ & $D^{\,(n)1}_{\phi,\rm NNL}$ &
 $D^{\,(n)2}_{\phi,\rm NNL}$ & $D^{\,(n)3}_{\phi,\rm NNL}$ \\[2mm] \hline
 & & & & & & \\[-2mm]
$1$  & 2.66667 & 1.33333 & 0       & 0       & 0       & 0\\[2mm]
$2$  & 14.5185 & 15.5556 & 0.888889& 104.051 & 8.59259 & 0\\[2mm]
$3$  & 41.5802 & 121.539 & 7.81893 & 897.918 & 79.0343 & 0.790123\\[2mm]
$4$  & 82.0448 & 489.196 & 31.2611 & 3947.46 & 378.981 & 6.23868\\[2mm]
$5$  & 123.863 & 1261.66 & 79.7128 & 11928.5 & 1228.05 & 25.3455\\[2mm]
$6$  & 151.299 & 2363.55 & 148.156 & 27287.3 & 2950.46 & 68.9743\\[2mm]
$7$  & 154.905 & 3455.21 & 215.475 & 49475.2 & 5530.83 & 139.478\\[2mm]
$8$  & 136.254 & 4124.46 & 256.338 & 73284.1 & 8381.24 & 222.015\\[2mm]
$9$  & 104.882 & 4147.64 & 257.174 & 90802.9 & 10551.2 & 289.008\\[2mm]
$10$ & 71.6724 & 3594.93 & 222.527 & 95974.4 & 11279.5 & 316.347\\[2mm]
$11$ & 43.9873 & 2732.85 & 168.950 & 87965.5 & 10424.5 & 297.446\\[2mm]
$12$ & 24.4770 & 1847.41 & 114.098 & 70891.3 & 8453.50 & 244.331\\[2mm]
\hline
\end{tabular}  
\vspace{2mm}
\caption{As Table 4, but for the coefficient function $C_{\phi,\rm q}$. 
\label{tab:CphiqNNLL}}   
\end{center}
\vspace*{-1mm}
\end{table} 

\begin{table}[tbh]
\vspace*{1mm}
\begin{center}
\small
\begin{tabular}{|c|ccccccc|}\hline
 & & & & & & & \\[-3mm]
 $n$ &
 $D^{\,(n)}_{L,\rm LL}$ & $D^{\,(n)1}_{L,\rm NL}$ &
 $D^{\,(n)2}_{L,\rm NL}$ & $D^{\,(n)1}_{L,\rm NNL}$ &
 $D^{\,(n)2}_{L,\rm NNL}$ & $D^{\,(n)3}_{L,\rm NNL}$ & 
 $D^{\,(n)4}_{L,\rm NNL}$ \\[2mm] \hline
 & & & & & & & \\[-2mm]
$1$ & 8      & 0      & 0      & 0      & 0      & 0      & 0\\[2mm]
$2$ & 48     & 160    & 0  & $-$35.0919 & 0      & 0      & 0\\[2mm]
$3$ & 144    & 1165.63& 10.6667& 3146.95& 103.111& 4.79341& 0\\[2mm]
$4$ & 288    & 4064.40& 64     & 24400.0& 974.683& 28.7605& 3.55556\\[2mm]
$5$ & 432    & 9222.47& 192    & 90273.6& 4214.99& 86.2814& 28.4444\\[2mm]
$6$ & 518.4  & 15447.8& 384    & 217656.& 11462.4& 172.563& 106.667\\[2mm]
$7$ & 518.4  & 20459.9& 576    & 387314.& 22429.4& 258.844& 256\\[2mm]
$8$ & 444.343& 22375.5& 691.2  & 544345.& 34032.4& 310.613& 448\\[2mm]
$9$ & 333.257& 20817.7& 691.2  & 630665.& 41989.1& 310.613& 614.4\\[2mm]
$10$& 222.171& 16840.7& 592.457& 620424.& 43532.4& 266.24 & 691.2\\[2mm]
$11$& 133.303& 12044.7& 444.343& 529638.& 38847.9& 199.68 & 658.286\\[2mm]
$12$& 72.7106& 7717.03& 296.229& 398936.& 30392.6& 133.12 & 543.086\\[2mm]
\hline
\end{tabular}  
\caption{The LL, NLL and NNLL coefficients of $C_{L,\rm g}$ in QCD,
as defined in Eq.~(\ref{eq:CLg-qcd}), to the 12-th order in $\ar=\as/(4\pi)$.
Also these coefficients further decrease at higher orders and tend to zero in
the infinite-order limit. The fifth column represents the $\flg11$ term 
absent in charged-current DIS.
\label{tab:CLgNNLL}}   
\end{center}
\end{table} 

%
{\footnotesize
\setlength{\baselineskip}{0.5cm}

}

\end{document}